\documentclass[acmsmall]{acmart}

\usepackage{colortbl}
\usepackage[most]{tcolorbox}
\usepackage[acronym]{glossaries}
\usepackage{makecell}
\usepackage{subcaption}
\usepackage{mdframed}
\definecolor{Gray}{gray}{0.9}

\AtBeginDocument{%
  \providecommand\BibTeX{{%
    \normalfont B\kern-0.5em{\scshape i\kern-0.25em b}\kern-0.8em\TeX}}}

\setcopyright{acmcopyright}
\copyrightyear{2021}
\acmYear{2021}

\acmBooktitle{Transactions on Software Engineering and Methodology (TOSEM)}

\newcommand{\etal}{~\textit{et al.}}

\begin{document}

\title{Broccoli: Bug localization with the help of text search engines}

\author{Benjamin Ledel}
\email{benjamin.ledel@tu-clausthal.de}
\affiliation{%
  \institution{TU Clausthal}
  \streetaddress{Arnold-Sommerfeld-Str. 1}
  \city{Clausthal-Zellerfeld}
  \state{Niedersachsen}
  \country{Germany}
  \postcode{38678}
}

\author{Steffen Herbold}
\email{steffen.herbold@tu-clausthal.de}
\affiliation{%
  \institution{TU Clausthal}
  \streetaddress{Arnold-Sommerfeld-Str. 1}
  \city{Clausthal-Zellerfeld}
  \state{Niedersachsen}
  \country{Germany}
  \postcode{38678}
}

\renewcommand{\shortauthors}{Benjamin Ledel et al.}

\begin{abstract}
    Bug localization is a tedious activity in the bug fixing process in which a software developer tries to locate bugs in the source code described in a bug report. Since this process is time-consuming and requires additional knowledge about the software project, information retrieval techniques can aid the bug localization process. In this paper, we investigate if normal text search engines can improve existing bug localization approaches. In a case study, we evaluate the performance of our search engine approach Broccoli against seven state-of-the-art bug localization algorithms on 82 open source projects in two data sets. Our results show that including a search engine can increase the performance of the bug localization and that it is a useful extension to existing approaches. As part of our analysis we also exposed a flaw in a commonly used benchmark strategy, i.e., that files of a single release are considered. To increase the number of detectable files, we mitigate this flaw by considering the state of the software repository at the time of the bug report. Our results show that using single releases may lead to an underestimation of the the prediction performance.
\end{abstract}


\newacronym{MAP}{MAP}{Mean Average Precision}
\newacronym{MRR}{MRR}{Mean Reciprocal Rank}
\newacronym{AST}{AST}{Abstract Syntax Tree}
\maketitle

\section{Introduction}
Maintaining a software project is a substantial exercise in the software developing process ~\cite{zhang2012empirical}. Due to its complexity, it is often time-consuming and therefore an expensive process ~\cite{marks2011studying, barbosa2019bulner}. For example, the Mozilla project receives hundreds of bug reports per day and each of them needs manual investigation ~\cite{thung2014buglocalizer}.

Bug localization is a tedious activity in which the software developer is searching for the part of the source code that contains the bug that is described in the bug report ~\cite{wang2016amalgamplus}. This is a difficult and possibly time consuming task, especially in large projects ~\cite{wang2016amalgamplus, thung2014buglocalizer, saha2013improving}. Since bugs only affect few files ~\cite{lucia2012faults}, Wang et al. ~\cite{wang2014version} compared the problem of localizing this small number of buggy files to search for the proverbial needle in a haystack.

There is a family of bug localization approaches (e.g., ~\cite{wong2014boosting, wang2016amalgamplus, saha2013improving}) that propose information retrieval techniques to support software developers. In these approaches, an algorithm computes a ranked list of files that potentially contain the bug described in the bug report ~\cite{lee}. The files are sorted by the likelihood that they contain the bug. 

However, a replication study by Lee et al. ~\cite{lee} reveals that the current approaches all exhibit a similar average performance on a larger data set. Furthermore, their results show that there is a large variance in the performance between different projects, which leads to an instability of the results. Lee et al.~\cite{lee} argue that despite recent efforts in bug localization, precision and recall values between 35 – 50\% may not be acceptable for a software developers. Additionally, recent research also mentioned major weaknesses in the way information retrieval techniques are evaluated, e.g., because of noise in the data ~\cite{kochhar2014potential}. However, Mills et al. ~\cite{mills2020relationship} show that it is possible to locate bugs with an appropriate score using only the bug report. They stress the importance of the query formulation for locating the bug. 

An aspect that is currently ignored by the bug localization literature are off-the-shelf search engines. Such search engines have been successfully applied to different domains, for example Semantic web ~\cite{lei2006semsearch}, bio-informatics ~\cite{letunic2012smart}, and 3D modeling ~\cite{funkhouser2003search}. Search engines use information retrieval to automatically optimize queries to achieve high-quality search results. They connect different techniques of information retrieval and are able to analyze query based data sets ~\cite{bialecki2012apache}. General search engines like Apache Lucene ~\cite{bialecki2012apache} and Elasticsearch\footnote{https://www.elastic.co/} are not developed for a particular domain, which indicates that they may also be suitable for bug localization. 

Considering the lack of performance and stability of current bug localization techniques on the one hand, and the successful application of search engines in different domains on the other hand, we propose our new bug localization algorithm \emph{Broccoli}, which combines the commonly used techniques for bug localization with an off-the-shelf search engine. One important property of search engines is that they internally reformulate queries for efficiency, which may be valuable for bug localization following the results of Mills et al.~\cite{mills2020relationship}. In order to explore the impact of search engines, our study answers the following research question:

\begin{itemize}
\item \textbf{RQ1:} How good are general search engines for bug localization? 
\end{itemize}

Based on the expectations from the literature, we derived the following hypothesis with respect to our research question: 
\begin{itemize}
\item \textbf{HT1:} Search engines are able to improve the mean performance of bug localization.
\end{itemize}

We derived \textbf{HT1} from the successful application of search engines in many fields (e.g.,~\cite{lei2006semsearch, letunic2012smart, funkhouser2003search}). Search engines were developed by a community that is dedicated to information retrieval and they include years of incremental improvements into their implementation. From our perspective, this is similar to relational databases: while other databases are better for certain use cases, the years of optimizations that went into relational databases are hard to outperform. We study HT1 through the definition of our novel bug localization approach called \textit{Broccoli} which combines existing bug localization techniques with an search engine. We accept HT1 if \textit{Broccoli} significantly outperforms existing bug localization approaches. 

Through our work, we identified an issue within the benchmarking strategy by Lee et al.~\cite{lee} that could affect the results of experiments. When replicating the benchmark, we found that often the bug locations could not possibly be found. A closer investigation revealed that this was due to the decision to use only the state of the software at major revisions to generate the ranking of potential bug locations. If file names changed between the major revision and the reporting of the bug, finding this bug is impossible. Thus, the benchmark is not aware of the state of the software at the time of reporting, to which we refer in the following as \textit{time-aware} approach. This finding led to our second research question.

\begin{itemize}
\item \textbf{RQ2:} How does the usage of major releases affect the evaluation of bug localization approaches in comparison to a time-aware approach?
\end{itemize}

Based on our expectations, we derived the following hypothesis regarding this research question.

\begin{itemize}
\item \textbf{HT2:} A time-aware approach can reduce the number of impossible to locate bugs which leads to a higher mean performance of bug localization approaches.
\end{itemize}

We derived \textbf{HT2} from the expectation that a time-aware bug localization that uses the state of the repository at the time a bug is reported not only provides a more realistic use case, but should also be more reliable when it comes to containing matching the bug descriptions to the files where the bugs are fixed, as they may have changed since a release. Only few cases should remain where localizing the bug is not possible, e.g., when the bug is not fixed on the main development branch and the other branch has different names, or when the file name changed between the reporting of the bug and the bug fix. If our assumptions that finding more bugs is possible, this should directly translate to an improvement of the performance values. Consequently, we accept \textbf{HT2} if we observe a significant difference when we compare the major release to the time-aware benchmark. 

Through our study of these research questions, we provide the following contributions to the state-of-the-art. 
\begin{itemize}
\item We extend the existing bug localization techniques with a search engine and found that this improves the results and that the search engine is the most important component among all aspects from the literature that were considered for bug localization.
\item We found that a time-aware bug localization leads to significantly different results such that benchmarks that do not account for the actual time of the bug report underestimate the performance of bug localization approaches. 
\end{itemize}

The remainder of this paper is structured as follows. In the Section~\ref{sec:related-work}, we summarize prior work on bug localization. Then, we present present our bug localization approach \textit{Broccoli} in Section~\ref{chapter:approach}. In Section~\ref{sec:experiments}, we present the experimental evaluation of our approach. In Section~\ref{sec:discussion}, discuss the implications regarding our research questions in Section~\ref{sec:discussion}. Afterwards, we discuss the threats to the validity of our results in Section~\ref{sec:threats} before we conclude in Section~\ref{chap:ende}.

\newpage

\section{Related work}
 
\label{sec:related-work}
\begin{table}[]
\centering
\footnotesize
\begin{tabular}{lp{11cm}}
{\textbf{Publication}} & {\textbf{Short description}} \\
\hline\hline
TFIDF-DHbPd ~\cite{sisman2012incorporating} & The first approach for bug localization used a simple matching based on term frequencies. While this work pioneered bug localization, the performance of this simple approach was consistently worst in the related work. Thus, we will not considered this in our experiments. \\ \hline
BugLocator ~\cite{zhou2012should} &
Zhou et al. ~\cite{zhou2012should} compute a vector space model from the source code in their approach BugLocator. The location of a bug is calculated from two scores: 1) the score from a query against the revised vector space model and 2) the  similarity to previous bug reports.
\\ \hline
BLUiR+ ~\cite{saha2013improving} &   
The approach uses an abstract syntax tree to extract source code terms such as classes, methods, and variable names. Then, it employs an open source information retrieval toolkit to find potential bug locations. The authors further extend this with the detection of similar bug reports from BugLocator.
\\  \hline
BRTracer+ ~\cite{wong2014boosting} &
Based on the idea that bug reports often contain stack traces, the authors of BRTracer+ divide each source code file into a series of segments and try to match these to the stack trace. This is combined with the vector space model and a parameterized version of BugLocator, that accounts for the different lengths of bug reports.
\\ \hline
AmaLgam+ ~\cite{wang2016amalgamplus} & 
Wang et al. combined the following five information retrieval techniques in their approach AmaLgam+: \emph{version history}, \emph{similar reports}, \emph{structure}, \emph{stack traces} and \emph{reporter information} ~\cite{wang2016amalgamplus}. The version history component is based on the BugCache algorithm by Rahman et al. ~\cite{rahman2011bugcache}. The similar report component is adapted from BugLocator ~\cite{zhou2012should}. Moreover, AmaLgam+ uses the structure component of BLUiR+ ~\cite{saha2013improving} to find package names, class names or method declarations. Similarly, AmaLgam+ detects stack traces in bug reports using the approach of BLUiR+ ~\cite{saha2013improving}. 
\\  \hline
BLIA ~\cite{youm2015bug} & 
Youm et al. ~\cite{youm2015bug} build their approach based on vector space model of BugLocator ~\cite{zhou2012should}. Moreover, BLIA tries to find similar bug reports from the bug repository. For the calculation of the similarity Youm et al. ~\cite{youm2015bug} use the bug comments of older reports as well as the bug summary and description. Then, the similarity score is computed via the cosine similarity. Additionally, BLIA analyzes the bug report for stack traces with the algorithm of BRTracer+ ~\cite{youm2015bug}. In the last step of the analytic process, Youm et al. ~\cite{youm2015bug} compute a score from the commit logs of the version history. Finally, all scores of the previous steps are combined by a linear expression with predefined control parameters.                   
\\  \hline
Locus ~\cite{wen2016locus} & 
Wen et al. proposed Locus, which offers a finer granularity in bug localization. Internally, Locus creates two corpora to predict potential bug locations: NL (Natural Language) and CE (Code Entities). The NL corpus contains the natural language tokens extracted from the selected hunks. In contrast, the CE corpus contains only package names, class names and method names. To calculate the ranked list of source code files, Locus constructs two queries for the two corpora (NL and CE). The final score is then computed using a linear combination of the two query results.          
\\  \hline
Blizzard ~\cite{rahman2018improving} &
Rahman et al. proposed a modified technique for bug localization that performs a query reformulation. The goal was to mitigate the noise in bug reports that inhibit the natural language processing. Blizzard determines whether there are excessive program entities or not in a bug report (query), and then applies appropriate reformulations to the query for bug localization. 
\\ \hline
\end{tabular}
\caption{Overview of the state-of-the-art bug localization approaches.}
\label{table:overview_long}
\end{table}

Within this section, we summarize the prior work on bug localization. Table~\ref{table:overview_long} lists and describes the previously suggested bug localization approach from the literature. A notable strength of the bug localization literature is that new approaches are usually the extension of a prior approach with an additional aspect that is considered. BLUiR+ and BRTracer+ both extend BugLocator. AmaLgam+ reuses components from BugLocator and BLUiR+. BLIA  uses components from BugLocator and BRTracer+. Only the approaches Locus and Blizzard are mostly independent from the other approaches. In general, the information used for locating the bugs can be categorized into the following components according to the literature: 
\begin{itemize}
    \item \textit{Similar reports:} using the similarity between the text of a bug report and code files to identify possible locations. 
    \item \textit{Version history:} exploitation of hot spots within the codes, i.e., the observation that files that contained bugs in the past often also contain bugs in the future. 
    \item \textit{Structure:} the direct identification of file names that may be affected through exploiting the frequent patterns of file names. 
    \item \textit{Stack trace:} exploiting information from stack traces as source for the bug localization.
    \item \textit{Reporter information:} use the assumption that the same reporter will use the same functionality of the software to find the package of the current bug.
    \item \textit{Bug report comments:} finding earlier bug reports by comparing the current bug with prior bugs, including their discussion. The location of the prior bug fixes can then be used to locate the current bug. 
\end{itemize}
Table~\ref{table:overview} summarizes which of the past approaches uses which of these techniques and also how our approach Broccoli fits within this picture. Similar to the related work, we did not develop a new approach from scratch. Instead, we rather defined an approach that re-uses components from the state of the art and includes a search engine as an additional component. 

\begin{table}[b]
\centering
\footnotesize
\begin{tabular}{l|p{1.5cm}p{1.5cm}p{1.5cm}p{1.5cm}p{1.5cm}p{1.5cm}}
\toprule
& \textbf{Version history} & \textbf{Similar reports} & \textbf{Structure} & \textbf{Stack trace} & \textbf{Reporter information} & \textbf{Bug report comments} \\ 
\midrule
TFIDF-DHbPd & \checkmark  & & & & & \\ \hline
BugLocator  &                          & \checkmark                      &                    &                      &                   &            \\ \hline
BLUiR+       &                          & \checkmark                      & \checkmark                &                      &                     &          \\  \hline
BRTracer+   &                          & \checkmark                      &                    & \checkmark                  &                       &        \\ \hline
AmaLgam+ & \checkmark                      & \checkmark                      & \checkmark                & \checkmark                  & \checkmark                   &        \\ \hline
BLIA & \checkmark                      & \checkmark                      & \checkmark                & \checkmark                  &                          &  \checkmark    \\ \hline
Locus    & \checkmark                      &                       &                 &                   &                          &  \\ \hline
Blizzard & \checkmark                  &                       &    \checkmark             &      \checkmark             &                          &  \\ \hline
Broccoli & \checkmark & \checkmark & \checkmark & \checkmark & & \\ \hline
\end{tabular}%
\caption{Overview of the techniques used by the state-of-the-art bug localization approaches.}
\label{table:overview}
\end{table}

In addition to the bug localization approaches, there are also general studies on the feasibility of bug localization. Kochhar et al.~\cite{kochhar2014potential} studied potential issues that could affect bug localization studies. They found that the biggest risk for bug localization studies is that they may overestimate the usefulness because some bugs may be trivial to locate, because the exact bug location is already part of the bug report. Most bug localization approaches directly harness this information through the structure component. Thus, while this may be relevant for the overall evaluation of the usefulness of bug localization approaches, this should not have a strong effect on benchmarks. Moreover, while such bugs may be simple to find, we note that developers are likely to reject a bug localization approach that misses the simple cases, which means that they should also be respected in the evaluation. 

Another general study was conducted by Mills et al.~\cite{mills2018bug, mills2020relationship}, who investigated if optimally formulated queries are able to find the locations of bugs. Through this, they studied if the information from the bug report is in theory sufficient to find a bug, given the optimal query would be found. Their results demonstrate that the text from the bug report contains a sufficient amount of information. However, since the approach optimizes the queries using the knowledge on the bug location, their study does not provide an approach that is feasible for the practical application of bug localization, which is why we cannot directly utilize their work in our comparison regarding the usefulness of search engines. 

\section{Broccoli: Using text search engines for bug localization}
\label{chapter:approach}

\begin{figure}
  \centering
  \includegraphics[width=\textwidth]{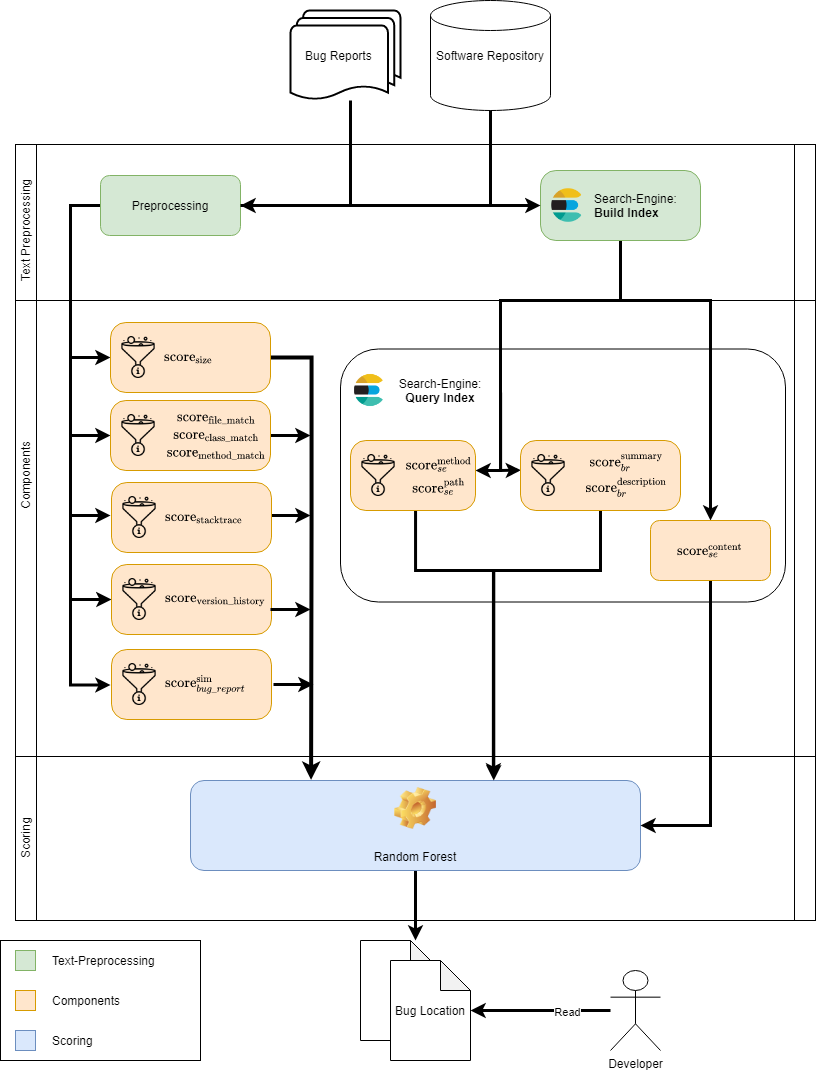}
   \caption{Design of Broccoli: Preprocessing steps are colored green, Components are colored orange and the combination as a final result is colored blue.}
	\label{fig:figure_approach_overview}
\end{figure}

In this chapter, we introduce our bug localization approach Broccoli, which can be considered as an extension of AmaLgam+ ~\cite{wang2016amalgamplus} with a text search engine. Figure \ref{fig:figure_approach_overview} shows the design of Broccoli. First, we preprocess the available data and prepare the index of the search engine.
Second, we use different approaches to determine the possible location of a bug, such as structural analysis, the identification of stack traces, source code risk analysis, and the search engine. Each of these components computes a score for every source file of the software repository.
Third, we combine these scores into a single score as final result. We use a random forest regression to aggregate the separate scores. The final output is a ranked list of bug locations that can be shown to developers.

\subsection{Text Preprocessing}
\label{section:preprocessing}

The goal of the preprocessing steps is to prepare the textual data of the bug reports and source code files for the analysis. This is done once at the beginning, because some components require the same preprocessing, e.g., the generation of an \gls{AST} of a source code file. We apply three different preprocessing methods to the data, that are then used by the components of Broccoli.

First, we create the \gls{AST} of all source code files. This enables components to easily extract entity names, such as class names and method names. Second, we apply a text processing steps to generate a harmonized representation of the bug descriptions and the source code files. 
\begin{itemize}
    \item We convert all characters to lower case. 
    \item We drop all non-alphanumeric characters. 
    \item We remove English stop words, i.e., words that occur very often and are, therefore, not suitable for localization. 
    \item We remove key words from the Java programming language. These can be considered as stop words within code, as they occur too often to be specific enough for localization. 
    \item We reduce the words to their word stem with a Porter stemmer~\cite{porter1980algorithm}. 
\end{itemize}
Third, we feed the issue reports, the source code files, and the entities from the \gls{AST} parsing into the search engine to prepare search indices.

\subsection{Components}
\label{sec:section_41}
Overall, Broccoli has seven scoring components that we describe in the following. 

\subsubsection{File size}
Lines Of Code (LOC) is a metric that represents the number of lines in a source code file. Since the literature on defect prediction (e.g., ~\cite{nagappan2005static}, ~\cite{shihab2013lines}) suggests that there is a correlation between the size of code files and the number of defects they contain, we can exploit this property for the localization of bugs. Thus, we define

\begin{equation}
\text{score}{_\text{size}}(f) = LOC(f)
\end{equation}
where $f$ is a file and $LOC(f)$ are the non-empty lines of the file. 

\subsubsection{Structure}
Our analysis of the structure is based on the approach suggested by Saha et al. ~\cite{saha2013improving} as part of BLUiR. Similar to BLUiR, we match based on file names, class names, and method names. We infer the identifiers from the source code through the generation of the \gls{AST}. We then search the bug summary and description for these identifiers to create three scores, i.e., one for matched file names, one for the class names, and one for the method declarations 

\begin{align}
\text{score}{_\text{file\_match}}(f, r) &= count(f_{name}, r) \\
\text{score}{_\text{class\_match}}(f, r) &= count(f_{classes}, r) \\
\text{score}{_\text{method\_match}}(f, r) &= count(f_{methods}, r)
\end{align}
where $f$ is a file, $r$ is a bug report, $f_{name}$ is the name of the file, $f_{classes}$ and $f_{methods}$ are the names of all classes and methods defined within $f$. The function $count$ evaluates how often the file name (with or without ending), respectively the defined classes and methods, occur within the bug report. 

\subsubsection{Stack trace}

Bug reports may contain a stack trace in case an application crashed as a result of an unhandled exception. The stack trace contains detailed data about the location in the source code where the unhandled exception was generated, including the method names, class names, and line numbers, as well as of the complete call stack. Since the stack trace is automatically generated, it contains reliable data that can be exploited for bug localization. However, the stack trace information is mixed with natural language. To detect stack traces in bug reports, we create an index, which contains all Java classes and file names. Then, we use a regular expression that matches phrases containing ``.java''. These phrases are checked against the file index. This removes all file names that are not in the code base (anymore). As a result, we have a list of files that may contain the bug. However, a stack trace reveals more information than just the Java classes, since the root cause of the bug may lie in some methods, which are used by the classes ~\cite{wong2014boosting}.
We follow Wong et al.~\cite{wong2014boosting} and add all directly mentioned source code files in the bug report to the set $D$. Then, we add all source code files from the import statement as well as all files from the same package to set $C$. However, the relevance for the bug localization differs between these sets ~\cite{wong2014boosting}. Schröter et al.~\cite{schroter2010stack} found that only the first 10 frames of a stack trace are relevant for bug localization. Therefore, the stack trace score is calculated as
\begin{equation}
\text{score}{_\text{stacktrace}}(f)=\left\{\begin{array}{ll}
\frac{1}{\operatorname{rank}(f)} & x \in D \land \operatorname{rank}(f) \leq 10 \\
0.1 & f \in C \land \operatorname{rank}(f)>10 \\
0.1 & f \in C \\
0 & \text {otherwise.}
\end{array}\right.
\end{equation}

where $f$ is a file, and $\operatorname{rank}(f)$ is the rank of the file within the stack trace ~\cite{wong2014boosting}. 

\subsubsection{Version history}

The version history component is similar to the version history component from AmaLgam+ ~\cite{wang2014version}, which is based on the FixCache algorithm by Rahman et al. ~\cite{rahman2011bugcache}. The algorithm takes the commit history of the source files as input and calculates a score for each file. Only commits that fulfill one of the following two criteria are considered.

\begin{itemize}
\item The commit message matches the regular expression: $(. * fix.*)|(. * bug.*)|(. * fail.*)|(. * error.*)$.
\item The commit was made in the last $k$ days before the reporting date of the issue. 
\end{itemize}

AmaLgam+ uses a fixed value of $k=15$. Since this value may be too low for projects with a low activity, we modified this condition: if there are less than 15 commits within the last 15 days, we increment $k$ until we have at least 15 commits.

The version history score of each source file $f$ is then calculated as 
\begin{equation}
\operatorname{score}_{version\_history}(f, k, C)=\sum_{c \in C \land f \in c} \frac{1}{\left.1+e^{12\left(1-\left(\left(k-t_{c}\right) / k\right) \right)}\right.}
\end{equation}
where $C$ is the relevant set of commits and $t_{c}$ is the number of days that have elapsed between a commit $c$ and the bug report. 

\subsubsection{Bug report similarity}
\label{section:bugreportsim}

Since similar errors is may be reported multiple times, it is useful to find similar and already fixed bug reports ~\cite{wang2016amalgamplus}. The similar bug report component assigns scores to files from fixed bug reports that are similar to the current bug report. We adopt the technique of BugLocator ~\cite{zhou2012should} and AmaLgam+ ~\cite{wang2016amalgamplus}. The bug report similarity considers all fixed bug reports that have been closed before the new bug report is submitted. We use the bug summary and description to find similar bug reports. Youm et al. ~\cite{youm2015bug} show that the comments of previous bug reports can improve the localization. Therefore, we add the comments to the corpus. Then, we compute a similarity score between each prior bug report $r'$ and the current report $r$ using $\operatorname{sim}\left(r, r'\right)$.

The score for source code file $f$, can then be computed as
\begin{equation}
\operatorname{score}_{bug\_report}^{\operatorname{sim}}(f, b, B)=\sum_{b \in \{b' \mid b' \in B \wedge f \in \operatorname{fix}(b')\}} \frac{\operatorname{sim}\left(b, b^{\prime}\right)}{|\operatorname{fix}(b')|}\label{eq:4}
\end{equation}
where $B$ is the set of previous bug reports from the database and $\operatorname{fix}(b^{\prime})$ is the set of files that are modified to fix the bug report $b^{\prime}$. The bug report similarity $\operatorname{sim}$ can be computed in several ways. We use two different measurements to define the $\operatorname{score}_{bug\_report}^{\operatorname{sim\_cos}}$ and $\operatorname{score}_{bug\_report}^{\operatorname{sim\_reporter}}$: cosine similarity, reporter similarity, and similarity using a search engine. The cosine similarity is defined as

\begin{equation}
\operatorname{sim}_{cos}(b, b')= \frac{\mathbf{b} \cdot \mathbf{b'}}{|\mathbf{b}|_{2}|\mathbf{b'}|_{2}} = \frac{\sum_{i=1}^{n} b_{i} \cdot b'_{i}}{\sqrt{\sum_{i=1}^{n} (b_{i})^{2}} \cdot \sqrt{\sum_{i=1}^{n} (b'_{i})^{2}}}
\end{equation}

where $\mathbf{b}, \mathbf{b'}$ are the vectors with the term frequencies of the bug reports. The reporter similarity

is defined as

\begin{equation}
\operatorname{sim}_{reporter}(b, b')= 
\begin{cases}
1 & \text{if}~b_{author} = b'_{author} \\
0 & \text{otherwise.}
\end{cases}
\end{equation}

This may help locating the bug, since a user may focus on using either one or a partial component of a software system, especially if the software system is large and provides several functionalities ~\cite{wang2016amalgamplus}.

\subsubsection{Search engine}

Search engines use indices to find relevant documents based on a search query. Internally, the search engine uses information retrieval techniques, including but not limited to stemming, lemmatization, and word embeddings ~\cite{bialecki2012apache}. The main contribution of our bug localization approach is the use of such a search engine as part of the bug localization process. 

We use the search engine in two ways. First, we utilize the search engine directly on the source code files. We create a document from each source code file. For each document, we generate three fields: the textual content of the source file, the path of the file, and the method names within the source file which are extracted as described in Section \ref{section:preprocessing}. We then conduct three queries to calculate scores. We calculate $score_{se}^{\operatorname{content}}$, $score_{se}^{\operatorname{method}}$, resp. $score_{se}^{\operatorname{path}}$ by querying the content, method, resp. path field with the preprocessed summary and description of the bug (see Section \ref{section:preprocessing}). The score $score_{se}^{\operatorname{path}}$ is only calculated if the bug summary or description contain the substring \textit{".java"} or \textit{"/"}, i.e., we find an indication for file names within the description. Otherwise, we set this score to 0. 

The $score_{se}^{\operatorname{content}}$ is the basic application of a search engine for bug localization, while the other two scores highlight semantic properties of the file that are often important, i.e., the contained methods and path of the file. 

The second way we use a search engine is to find similar bug reports. A search engine is able to present a more accurate result, since it goes beyond cosine similarity for text matching. We create a document for each bug report with three fields: the closing date, the summary, and the content. We query this search index twice: we use the bug summary to calculate $score_{br}^{\operatorname{summary}}$ and the bug description to calculate $score_{br}^{\operatorname{description}}$. The closing date field is used to define a condition that excludes all documents for bugs that were not fixed upon the time of the creation of the bug that should be localized. The query evaluates the summary and the content field at once, i.e., we let the search engine combine the results of the search on the two fields. We then use these similarities in the same way as described in Section \ref{section:bugreportsim} to calculate $\operatorname{score}_{bug\_report}^{\operatorname{sim}}$ according to Equation (\ref{eq:4}). 

\subsection{Scoring: Combination of components}

We described how different approaches can be used to create scores for each file that indicate their likelihood to contain the bug in Section \ref{sec:section_41}. We now combine these scores into a single score. The resulting score can then be used in descending order to sort files and thereby provide possible bug locations to the software developer.

In comparison to the related work that used linear combinations to combine scores ~\cite{wang2016amalgamplus, wong2014boosting}, we use a random forest regression to determine the final scores. Since Random Forest uses decision trees, it can also capture non-linear relationships ~\cite{breiman2001random}. The training data can be collected from past closed bug reports and their linked commits. All files that are modified in the linked commits have a 1 in the result column, 0 otherwise.

\section{Experiments}
\label{sec:experiments}

We now describe the experiments we conducted to compare the bug localization approaches and to explore how the different scores affect the results of Broccoli. The experiments are based on the benchmarking tool and data of Bench4BL~\cite{lee}. We extended this benchmark with a second data set and also evaluate how information leakage due to removed or renamed files affects the results. The complete code to conduct our experiments is provided as a replication package \footnote{https://github.com/benjaminLedel/broccoli\_replicationkit}. In the following, we describe the data, baselines for a comparison with Broccoli, performance criteria, evaluation methodology, and results. 

\subsection{Data}
We use two data sources to prepare into three data sets to conduct our experiments. The first data set was published by Lee et al. ~\cite{lee} as part of Bench4BL and consists of 51 open source projects written in Java from Apache Foundation\footnote{http://www.apache.org}, Spring\footnote{http://spring.io} and JBoss\footnote{http://www.jboss.org}. Table \ref{tab:Bench4BL_data} gives an overview of the data. We include all source files as well as the test files from the project.

\begin{table}[]
\footnotesize
\centering
\begin{tabular}{lllll}
\textbf{Group}                & \textbf{Project} & \textbf{\#Files (Max)} & \textbf{\#Versions} & \textbf{\#Bugs} \\ \hline \hline
Apache      & camel            & 14522                         & 60                        & 1419                   \\ 
            & hbase            & 2714                          & 70                        & 756                    \\  
            & hive             & 4651                          & 21                        & 971                    \\  
    
    & \textit{commons-codec}            & 115                           & 9                         & 40                     \\  
            & \textit{commons-collections}      & 525                           & 7                         & 76                     \\  
            & \textit{commons-configuration}    & 447                           & 11                        & 129                    \\  
            & commons-crypto           & 82                            & 1                         & 8                      \\  
            & commons-csv              & 29                            & 3                         & 14                     \\  
            & \textit{commons-io}              & 227                           & 13                        & 84                     \\  
            & \textit{commons-lang}             & 305                           & 16                        & 194                    \\  
            & \textit{commons-math}             & 1617                          & 15                        & 237                    \\  
            & weaver           & 113                           & 1                         & 2                      \\ \hline
JBoss       & entesb           & 252                           & 3                         & 46                     \\
            & jbmeta           & 858                           & 5                         & 25                     \\  
            & ely              & 68                            & 3                         & 24                     \\  
            & swam            & 727                           & 6                         & 57                     \\  
            & wfarq            & 126                           & 1                         & 1                      \\  
            & fcore           & 3598                          & 16                        & 353                    \\  
            & wlfy             & 8990                          & 11                        & 957                    \\  
            & wfmp             & 80                            & 1                         & 3                      \\ \hline
Spring      & amqp             & 408                           & 33                        & 105                    \\ 
            & android          & 305                           & 2                         & 11                     \\  
            & batch            & 1732                          & 33                        & 429                    \\  
            & batchadm         & 243                           & 4                         & 20                     \\  
            & datacmns         & 604                           & 33                        & 143                    \\  
            & datagraph        & 848                           & 4                         & 58                     \\  
            & datajpa          & 330                           & 38                        & 136                    \\  
            & datamongo        & 622                           & 40                        & 252                    \\  
            & dataredis        & 551                           & 17                        & 48                     \\  
            & ldap             & 566                           & 5                         & 52                     \\  
            & mobile           & 64                            & 3                         & 11                     \\  
            & roo              & 1109                          & 15                        & 642                    \\  
            & sec              & 1618                          & 42                        & 473                    \\  
            & secoauth         & 726                           & 7                         & 101                    \\  
            & sgf              & 695                           & 19                        & 106                    \\  
            & shdp             & 1102                          & 9                         & 45                     \\  
            & shl              & 151                           & 3                         & 10                     \\  
            & social           & 212                           & 4                         & 15                     \\  
            & socialfb         & 252                           & 5                         & 15                     \\  
            & socialli         & 180                           & 1                         & 4                      \\  
            & socialtw         & 153                           & 5                         & 8                      \\  
            & spr              & 6512                          & 12                        & 57                     \\  
            & swf              & 808                           & 20                        & 129                    \\  
            & sws              & 925                           & 25                        & 162                    \\ \hline
Old subjects& aspectj          & 6485                          & 1                         & 251                    \\  
            & jdt              & 6842                          & 1                         & 70                     \\  
            & pde              & 5273                          & 1                         & 40                     \\  
            & swt              & 484                           & 1                         & 41                     \\  
            & zxing            & 391                           & 1                         & 20                     \\ \hline \hline
\textbf{Total} &               & \textbf{80906}                         & \textbf{695}                       & \textbf{9074}                   \\ 
\end{tabular}
\caption{Overview of the Bench4BL data. Projects highlighted in italic are also part of the SmartSHARK data.}
\label{tab:Bench4BL_data}
\end{table}

The second data set is extracted from the SmartSHARK database ~\cite{trautsch2021smartshark}. The data set consists of a convenience sample of projects from the Apache Software Foundation.  We used all 38 projects with manually validated issue types to ensure that our data contains only bug reports ~\cite{herbold2020issues} this excludes feature request issues. Thus, this second data set avoids problems due to mislabeled issue types described by Kochhar et al.~\cite{kochhar2014potential}. Table \ref{tab:smartshark_data} gives an overview of the SmartSHARK data. There is an overlap of seven projects with the Bench4BL data, which we highlight in the table. However, the data for these projects are not identical, due to the different times of data collection and the manual validation by Herbold et al.~\cite{herbold2020issues}. We followed the approach from the Bench4BL data and used the latest release before the bug was reported to collect the files to which the bug reports could potentially be matched. 

\begin{table}[]
\footnotesize
\centering
\begin{tabular}{llll}
\textbf{Project}  & \textbf{\#Files (Max)} & \textbf{\#Versions} & \textbf{\#Bugs} \\ \hline\hline
ant-ivy           & 639                           & 6                         & 374                    \\ 
archiva           & 728                           & 7                         & 214                    \\ 
calcite           & 1748                          & 16                        & 306                    \\ 
cayenne           & 2779                          & 2                         & 242                    \\ 
commons-bcel      & 489                           & 6                         & 28                     \\ 
commons-beanutils & 245                           & 10                        & 24                     \\ 
 \textit{commons-codec}     & -                           & 11                          & 30            \\
 \textit{commons-collections}     & -                           & 9                          & 41                     \\
 \textit{commons-compress}     & -                           & 17                         & 106                     \\
 \textit{commons-configuration}     & -                           & 14                          & 123                     \\
commons-dbcp      & 120                           & 11                        & 65                    \\ 
commons-digester  & 308                           & 14                        & 6                      \\ 
 \textit{commons-io}     & -                           & 11                          & 45                     \\ 
commons-jcs       & 561                           & 6                         & 54                     \\ 
commons-jexl      & 141                           & 6                         & 45                     \\ 
 \textit{commons-lang}     & -                           & 16                          & 143                     \\ 
 \textit{commons-math}     & -                           & 13                          & 204                     \\ 
commons-net       & 404                           & 15                        & 117                    \\
commons-scxml     & 144                           & 5                         & 17                     \\
commons-validator & 147                           & 7                         & 45                     \\
commons-vfs       & 382                           & 4                         & 84                    \\
deltaspike        & 1727                          & 16                        & 118                    \\
eagle             & 1853                          & 3                         & 100                    \\
giraph            & 1014                          & 3                         & 109                    \\
gora              & 442                           & 8                         & 51                     \\
jspwiki           & 998                           & 12                        & 119                    \\
knox              & 1050                          & 13                        & 124                    \\
kylin             & 1272                          & 11                        & 279                    \\
lens              & 856                           & 2                         & 184                    \\
mahout            & 1376                          & 13                        & 222                    \\
manifoldcf        & 1305                          & 28                        & 262                    \\
nutch             & 620                           & 22                        & 376                    \\
opennlp           & 842                           & 2                         & 87                    \\
parquet-mr        & 630                           & 10                        & 49                     \\
santuario-java    & 660                           & 6                         & 47                     \\
systemml          & 1749                          & 10                        & 258                    \\
tika              & 1000                          & 28                        & 320                    \\
wss4j             & 719                           & 5                         & 150                    \\
\hline\hline
\textbf{Total}    & \textbf{26948}                         & \textbf{291}                       & \textbf{5168}                   \\
\end{tabular}
\caption{Overview of the SmartSHARK data. Projects highlighted in italic are also part of the Bench4BL data.}
\label{tab:smartshark_data}
\end{table} 

The third data set contains all projects from the SmartSHARK data source. In addition to the second data set, this data set also includes the italic projects. In our experiments we use this data set to investigate the influence of different matching strategies. Thus, we apply our time-aware matching strategy, because the major release matching as used in the state-of-art has some disadvantages. In a practical application, the date of the a release usually does not match the date of the bug report creation. Moreover, the bug could even refer to a yet unreleased version of a project. In case new behavior was added, e.g., new GUI components, the bug description may not make sense if the latest release is matched based on the reporting date. However, this is the case in the matching strategy of Lee et al.~\cite{lee}, who suggest to use the code base at the time of the latest release. To mitigate this issue, we propose the following time-Aware matching rules: 
\begin{enumerate}
\item[1.] Check the availability of the fixed versions field of the bug report:
\begin{description}
\item[a.] if the field is available, execute the bug localization on each of the fixed versions. 
\item[b.] otherwise, execute the bug localization against all prior versions that contain the modified files. 
\end{description} 
\item[2.] Calculate the score as
\begin{equation}
score(report) = \sum_{i = 1}^{n}\frac{score_{\text{timeaware}}(i, report)}{n}
\end{equation}
where $report$ is the bug report under investigation, $score_{\text{timeaware}}(i, report)$ is the $i$-th nearest subscore of the bug localization algorithm, and $n$ is the total number of version for the bug report.
\end{enumerate} 

\subsection{Baselines}

We use the state-of-the-art of bug localization as baselines for the evaluation of Brocolli. Thus, we compare Brocolli to BugLocator ~\cite{zhou2012should}, BLUiR ~\cite{saha2013improving}, BRTracer+ ~\cite{wong2014boosting}, AmaLgam ~\cite{wang2014version}, BLIA  ~\cite{youm2015bug}, Locus ~\cite{wen2016locus} and Blizzard ~\cite{rahman2018improving}. Where possible, we re-use the implementations from Bench4BL to avoid potential issues due to implementation errors. For BLUiR+ and AmaLgam+ we extended the implementation of Bench4BL with the bug report similarity, as Bench4BL contained only BLUiR and AmaLgam, which do not use these components. Because the publications do not indicate how the scores from these components are weighted, we assume equal weighting. Since Blizzard is not available in Bench4BL, we re-use the implementation the authors provided as part of their replication kit~\cite{masud_rahman_2018_1297907}. We use the parameters specified in the original work of the bug localization techniques. 

\subsection{Performance metrics}
We use two performance metrics to evaluate the quality of the estimated bug locations, again following the the Bench4BL benchmark~\cite{lee}.

The \textit{Mean Average Precision} (\gls{MAP}) considers the ranks of all buggy files ~\cite{lee}. Therefore, \gls{MAP} prioritizes recall over precision and is mostly relevant in scenarios where the software developer analyzes the whole ranked list to find many relevant results or buggy files ~\cite{wang2016amalgamplus}. The average precision of a single query is computed as:
\begin{equation}A P=\sum_{k=1}^{M} \frac{P(k) \times p o s(k)}{\text {number of positive instances}}
\end{equation}
$M$ is the number of ranked files, $k$ is the rank in the ranked file list. $Pos(k)$ is 1 if the $k-th$ file is included in the fixed file list, otherwise $pos(k)$ is 0. $P(k)$ is the precision and computed as: 
\begin{equation}
P(k)=\frac{\text {number of buggy files }}{k}.
\end{equation}
Using the average precision, \gls{MAP} can be computed as: 
\begin{equation}
M A P=\frac{1}{M} \sum_{j=1}^{M} A P(j).
\end{equation}

The \textit{Mean Reciprocal Rank} (\gls{MRR}) considers the ranks of the  first buggy file. The rank of this file is called the reciprocal rank for that query ~\cite{wang2016amalgamplus}. \gls{MRR} is the mean of the reciprocal ranks over all queries $Q$ and is computed as:
\begin{equation}
M R R=\frac{1}{|Q|} \sum_{i=1}^{|Q|} \frac{1}{\operatorname{rank}_{i}}
\end{equation}
The \gls{MRR} value is based one the principle that a software developer will look at each hit until the first relevant document appears ~\cite{lee}. The \gls{MRR} value is identical to \gls{MAP} in cases where each query has exactly one relevant buggy file ~\cite{Craswell2009}.

\subsection{Methodology}
Our general methodology for the experiments consists of three phases. In Phase 1, we conduct a leave-one-project-out cross validation experiment with the Bench4BL data. For this, we use one project as test data and all other projects as training data. We then run the bug localization with Brocolli and our baselines and measure the \gls{MAP} and \gls{MRR}. Thus, the first phase is an replication of the Bench4BL benchmark, with the addition of Brocolli. 

In Phase 2, we use the Bench4BL data to train a prediction model and evaluate how well this model generalizes to other data on the SmartSHARK data without the projects that are already part of Bench4BL. This design allows us to evaluate the performance of bug localization in a realistic scenario, where a model trained by a vendor is deployed for bug localization within products without local re-training or data collection at the customer site. 

In Phase 3, we use the time-aware variant of the SmartSHARK data. Through this, we evaluate the performance of the bug localization in a more realistic setting. The comparison of these results with the results from Phase 2 allows us to see if the additional effort to create the time-aware data is relevant for benchmarking bug localization approaches, or if the version matching is sufficient.

In all three phases, we choose for Broccoli to train Random Forest with 1000 trees. The number of trees is selected to find an reasonable tradeoff between the training time and the prediction performance ~\cite{oshiro2012many}.
We evaluate the results of all three phases following the guidelines for the comparison of classifiers by Benavoli\etal~\cite{benavoli2017time} as implemented in the Autorank package~\cite{Herbold2020}. These guidelines are an updated version of the popular guidelines by Demsar~\cite{demvsar2006statistical}. In comparison to the original guidelines, the authors suggest to use a Bayesian approach for the statistical analysis instead of a frequentist approach. Following the guidelines by Benavoli\etal~\cite{benavoli2017time}, we use the Bayesian signed rank test~\cite{benavoli2014bayesian} with a uniform prior. We use the Shapiro-Wilk test~\cite{razali2011power} to determine if the data is normal. For normal data, we follow Kruschke and Liddell~\cite{kruschke2018bayesian}, we set the Region of Practical Equivalence (ROPE) as $\pm 0.1 \cdot d$, where $d$ is the effect size according to Cohen (Cohen's $d$)~\cite{cohen2013statistical}. For two populations $A$ and $B$, the effect size $d$ is defined as
\begin{equation}
d = \frac{mean(A)-mean(B)}{s}
\end{equation}
with $s$ is the pooled standard deviation of population
\begin{equation}
s = \sqrt{\frac{(n_A-1) \cdot std(A)^2+(n_B-1) \cdot std(B)^2}{n_A+n_B-2}}
\end{equation}
and $n_A$ and $n_B$ the number samples of the populations A and B. In our case the populations are the results for a bug localization approach and the sample size the number of projects to which we apply the approach. For brevity, we omit how Autorank treats non-normal data.\footnote{As reported later, all our results are normal.}

The ROPE defines a region around the mean value in which differences are considered so small, that they would not have a practical impact. Setting the rope to $\pm 0.1 \cdot d$ means that the differences in values is not even as large a half the size of a small effect are considered to have no practical effect and are, therefore, considered as \emph{practically equivalent}. For two approaches $A$ and $B$, the Bayesian signed rank test determines the posterior probabilities $p_{A}, p_{equal}$, and $p_{B}$ they describe the probability that $A$ is larger, the populations are practically equivalent, or that $B$ is larger. Following Benanvoli\etal~\cite{benavoli2017time}, we accept the hypothesis that a population $A$ is larger/equal/smaller than population $B$ if $p_{A} \geq 0.95$/$p_{equal} \geq 0.95$/$p_{B} \geq 0.95$. If none of the three probabilities is larger than $0.95$, the result of the comparison is inconclusive. 
We apply the Bayesian signed rank test to all pairs of approaches in order to determine a ranking. Report the mean value, standard deviation, and confidence interval of the mean value for all populations. We use Bonferroni correction~\cite{napierala2012bonferroni} for the calculation of the Shapiro-Wilk test and the confidence intervals to ensure a family-wise confidence level of 95\%. Additionally, we rank the populations by the mean value and report the effect size $d$ in comparison to the best ranked population, as well as the probability of being practically equivalent or smaller than the best ranked population, and the decision made based on the posterior probabilities. 

Orthogonal to the analysis of the performance of the different approaches, we also evaluate which of the various scores we compute are relevant for the bug localization. We utilize the feature importance of the Random Forest we train for aggregating the scores of Broccoli into the final result. The feature importance measures how often each feature is used by the Random Forest to make decisions~\cite{biau2016random}. We compute the feature importance for all Random Forests during all phases of the experiments and compare them visually using box plots. In addition to learning which scores are important for the bug localization, we can also analyze the stability of these results to increase the validity of our findings. The comparison of the feature importance between the Bench4BL data and the SmartSHARK data allows us to determine if the results are stable across data sets collected by different research groups. The comparison of the results between the version-matching and the time-aware results on the SmartSHARK data allows us to determine if certain features may only be important due to information leakage. We conduct this analysis using Broccoli because Broccoli can be considered as a superset in terms of used scores in comparison to the other approaches.

\subsection{Results}
In this section, we present the results of our empirical study. 

\subsubsection{Results for Phase 1}

\begin{table}
\begin{tabular}{lrrrrrr}
\textbf{Approach} & \textbf{Mean MAP} & \textbf{Best MAP} & $d$ & \textbf{Mean MRR} & \textbf{Best MRR} & $d$\\
\hline\hline
Broccoli & 0.459$\pm$0.103 & 23 & - &  0.523$\pm$0.102 & 20 & - \\
Locus & 0.426$\pm$0.091 & 10 & 0.189 & 0.490$\pm$0.093 & 6 & 0.100 \\
AmaLgam+ & 0.424$\pm$0.109 & 3 & \textbf{0.188} & 0.482$\pm$0.103 & 3.83 & \textbf{0.165} \\
BLUiR+ & 0.417$\pm$0.110 & 4 & \textbf{0.223} & 0.476$\pm$0.105 &  3.83 & \textbf{0.187} \\
BRTracer+ & 0.412$\pm$0.087 & 4 & \textbf{0.281} & 0.505$\pm$0.095 & 6.5 & 0.216 \\
BugLocator & 0.399$\pm$0.084 & 2 & \textbf{0.363} & 0.486$\pm$0.090 & 3.5 & 0.222 \\
BLIA & 0.394$\pm$0.092 & 5 & \textbf{0.373} & 0.492$\pm$0.105 & 8.33 & 0.255 \\
Blizzard & 0.259$\pm$0.058 & - & \textbf{1.346} & 0.256$\pm$0.057 & - & \textbf{1.820} \\
\hline
\end{tabular}
\caption{Results for the Bench4BL data. We report the mean performance, confidence intervals (as $\pm$), the number of times with the best performance of the approaches, and the effect size in comparison to Broccoli. A bold-faced effect size indicates that the difference is significant. BLUiR+ and AmaLgam+ share five first places for \gls{MAP} and \gls{MRR}, which we count as 0.5 for both. BLUiR+, AmaLgam+ and BLIA share one first place for \gls{MRR}, which we count as 0.33 for each.}
\label{tbl:phase1}
\end{table}

\begin{figure}
\centering
\begin{subfigure}[b]{0.48\textwidth}
\centering
\includegraphics[width=\textwidth]{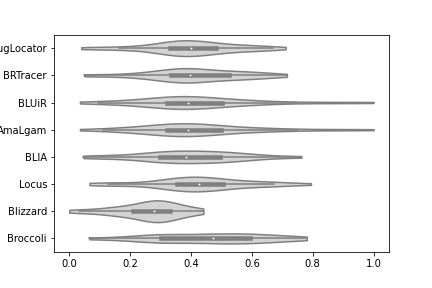}
\caption{MAP values}
\label{fig:experiment1map}
\end{subfigure}
\begin{subfigure}[b]{0.48\textwidth}
\centering
\includegraphics[width=\textwidth]{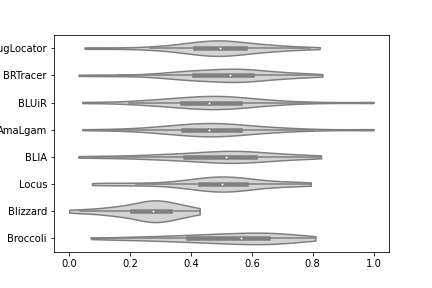}
\caption{MRR values}
\label{fig:experiment1mrr}
\end{subfigure}
\caption{Violin plots of the MAP and MRR values on the Bench4BL data.}
\label{fig:experiment1}
\end{figure}

Table \ref{tbl:phase1} and Figure~\ref{fig:experiment1} summarize the results of the first phase, i.e., our replication of the Bench4BL benchmark protocol. We observe that the overall performance is relatively uncertain, i.e., the confidence intervals of the mean performance are relatively large the violin plots \ref{fig:experiment1} further demonstrate the spread of the results. Regardless, Broccoli achieves the best mean \gls{MAP}. and \gls{MRR}. The difference to Locus is not significant in either metric. However, the Bayesian signed ranked tests indicates that there is a probability of about 90\% that the mean value of Broccoli is larger than Locus. Thus, while not (yet) significant, this provides a strong indication that this is only due to the large standard deviation and the small amount of data. For all other approaches, Broccoli achieves significantly higher \gls{MAP}. With \gls{MRR}, the difference to BRTracer+, BugLocator, and BLIA is not significant. We observe that, with the exception of Blizzard, the effect sizes are relatively small. Additionally, we observe that Broccoli is the best approach for about half of the projects, while the other approaches only sporadically rank first. This indicates that while there is also a high uncertainty in the results of Broccoli, there seems to be a small but consistent mean improvement with Broccoli and that Broccoli mostly does not rank first when other approaches have a positive outlier. Detailed results for each project are in Appendix~\ref{sec:details}.

\subsubsection{Results for Phase 2}

\begin{table}
\begin{tabular}{lrrrrrr}
\textbf{Approach} & \textbf{Mean MAP} & \textbf{Best MAP} & $d$ & \textbf{Mean MRR} & \textbf{Best MRR} & $d$\\
\hline\hline
Broccoli & 0.474$\pm$0.253 & 16 & - &  0.517$\pm$0.116 & 4 & 0.172 \\
Blizzard & 0.444$\pm$0.217 & 5 & \textbf{0.219} & 0.418$\pm$0.098 & 1 & \textbf{0.909} \\
BRTracer+ & 0.442$\pm$0.271 & 3 & \textbf{0.251} & 0.542$\pm$0.111 & 16 & - \\
BLIA & 0.418$\pm$0.264 & 5 & \textbf{0.434} & 0.513$\pm$0.118 & 8 & \textbf{0.195} \\
BLUiR+ & 0.402$\pm$0.251 & 2 & \textbf{0.523} & 0.390$\pm$0.107 &  - & \textbf{1.070} \\
BugLocator & 0.388$\pm$0.276 & - & \textbf{0.679} & 0.505$\pm$0.109 & 2 & \textbf{0.260} \\
Locus & 0.261$\pm$0.273 & - & \textbf{2.020} & 0.337$\pm$0.111 & - & \textbf{1.413} \\
AmaLgam+ & 0.097$\pm$0.174 & - & \textbf{3.803} & 0.164$\pm$0.066 & - & \textbf{3.167} \\
\hline
\end{tabular}
\caption{Results for the SmartSHARK data. We report the mean performance, confidence intervals (as $\pm$), the number of times with the best performance of the approaches, and the effect size in comparison to Broccoli. A bold-faced effect size indicates that the difference is significant. }
\label{tbl:phase2}
\end{table}

\begin{figure}
\centering
\begin{subfigure}[b]{0.48\textwidth}
\centering
\includegraphics[width=\textwidth]{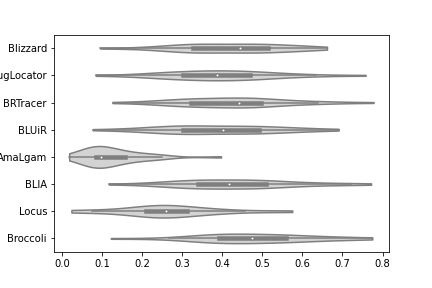}
\caption{MAP values}
\label{fig:experiment2map}
\end{subfigure}
\begin{subfigure}[b]{0.48\textwidth}
\centering
\includegraphics[width=\textwidth]{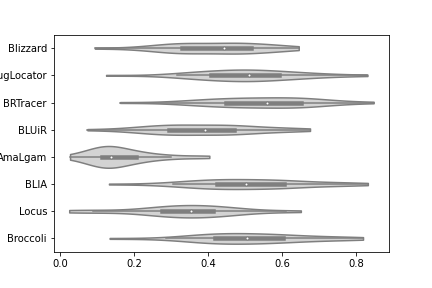}
\caption{MRR values}
\label{fig:experiment2mrr}
\end{subfigure}
\caption{Violin plots of the MAP and MRR values on the SmartSHARK data.}
\label{fig:experiment2}
\end{figure}

Table~\ref{tbl:phase2} and Figure~\ref{fig:experiment2} summarize the results of the second phase, i.e., the generalization of a bug localization model that was created using Bench4BL applied to the SmartSHARK data. Similar to the first phase, all performance estimates have a high variance and Broccoli achieves the best mean \gls{MAP}. With \gls{MAP}, Broccoli dominates the other approaches with the best result for 16 of the 31 projects. The difference to the other approaches is significant. The results for \gls{MRR} diverge from the results on the Bench4BL data, due to a very good performance by BRTracer+, which has the the best mean performance and yields the best overall result on 16 projects. BLIA is relatively similar to Broccoli, with a slightly lower mean performance, but higher ranks results. Both BRTracer+ and BLIA are not significantly different from Broccoli. Overall, this indicates that Broccoli is consistently and significantly better with a medium effect with respect to MAP and that there is a no improvement over the state of the art with respect to \gls{MRR}. 

\subsubsection{Results for Phase 3}

In the third phase of the experiment, we evaluate the impact of time-awareness to evaluate if the possible noise in the data due to the attempt to determine the number of impossible matches. Table \ref{tab:statistics} summarizes how the data is affected. Overall, there are 18,690 files that are modified for the correction of 5,190 bugs. With the approach to use the state of the project at major releases, as suggested by Bench4BL and as we used so far, only 14.830 files are detectable, for the remaining 4,160 the localization is impossible, because the files that were corrected did not exist at the time when the data was collected during the last major release. For 562 bug reports, none of the affected files can be detected. Thus, for about 10.8\% of the bugs the localization is impossible with this approach. 

This changes with the time-aware strategy: the number of undetectable files drops from 4,160 to 2,556, that means that about 61\% of the undetectable files can now be found within the repository. Due to this improvement, there are only 130 bugs left, for which none of the affected files can be detected. Hence, there is a reduction from 10.8\% to 0.3\% of the impossible to located bugs with the time-aware approach.

\begin{table}[!b]
\centering
\begin{tabular}{lcc}
      & \textbf{Multiversion approach} & \textbf{Time-aware approach} \\
\hline
Files & 4,160 (22.6\%)  & 2,556 (13.7\%) \\
Bugs  & 562 (10.8\%)      & 130 (0.3\%) \\
\hline
\end{tabular}
\caption{Comparison between using major releases and a time aware approach. The table lists the undetectable files and bugs versus the total number of files and bugs. Our data set contains 18,690 files and 5,190 bug reports.}
\label{tab:statistics}
\end{table}

\begin{table}
\begin{tabular}{lrrrrrr}
\textbf{Approach} & \textbf{Mean MAP} & \textbf{Best MAP} & $d$ & \textbf{Mean MRR} & \textbf{Best MRR} & $d$ \\
\hline\hline
Time-aware & 0.587$\pm$0.067 & 33 & - & 0.634$\pm$0.074 & 35 & - \\
Release approach & 0.505$\pm$0.080 & 8 & \textbf{0.559} & 0.546$\pm$0.089 & 6 & \textbf{0.569} \\
\hline
\end{tabular}
\caption{Comparison between the results of Broccoli on the version and the time-aware SmartSHARK data. We report the mean performance, confidence intervals (as $\pm$), the number of times with the best performance of the approaches, and the effect size in comparison to Broccoli on the time-aware data. A bold-faced effect size indicates that the difference is significant.}
\label{tbl:phase3}
\end{table}

\begin{figure}
\centering
\begin{subfigure}[b]{0.48\textwidth}
\centering
\includegraphics[width=\textwidth]{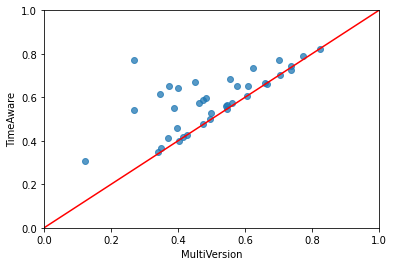}
\caption{MAP values}
\label{fig:experiment3map}
\end{subfigure}
\begin{subfigure}[b]{0.48\textwidth}
\centering
\includegraphics[width=\textwidth]{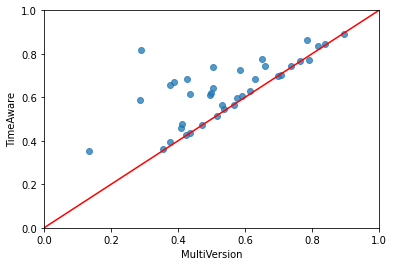}
\caption{MRR values}
\label{fig:experiment3mrr}
\end{subfigure}
\caption{Scatterplot that compares the results of the version and time-aware approach. Each point represents the scores of one project. The diagonal indicates equal performance.}
\label{fig:experiment3}
\end{figure}

Table~\ref{tbl:phase3} summarizes how the evaluation changes with the more accurate time-aware data. The results show that the time-aware performance is significantly different: the performance is significantly higher with a effect size of about 0.56 for both MAP and MRR. Figure~\ref{fig:experiment3} provides a detailed view of the deviations. We observe that while the performance estimate is sometimes accurate (on the diagonal), there are also many cases where the release approach underestimates the performance and there are no cases where the performance is overestimated by the release approach.

\subsubsection{Results for the Feature Importance}

\begin{figure}
\centering
\begin{subfigure}[b]{0.48\textwidth}
\centering
\includegraphics[width=\textwidth]{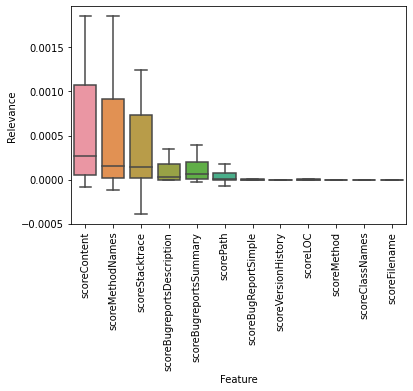}
\caption{Feature frequency in Phase 1}
\label{fig:featureImportancePhase1}
\end{subfigure}
\begin{subfigure}[b]{0.48\textwidth}
\centering
\includegraphics[width=\textwidth]{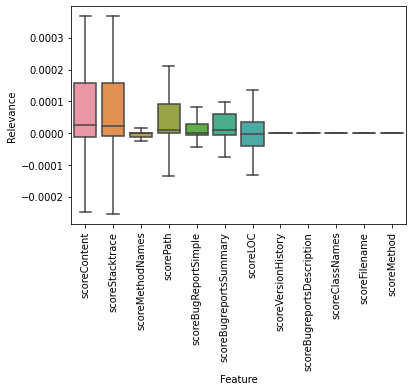}
\caption{Feature frequency in Phase 2}
\label{fig:featureImportancePhase2}
\end{subfigure}
\caption{Feature frequency of the random forest model in Phase 1 and Phase 2.}
\label{fig:featureImportance}
\end{figure}

Figure \ref{fig:featureImportance} illustrates the feature permutation importance of the Random Forest used by Broccoli. We observe that there is almost no difference between the two phases, i.e., the results are stable across data sets. The feature  $score_{se}^{\operatorname{content}}$ is the most relevant feature in both. The $\text{score}{_\text{stacktrace}}$ and $\text{score}{_\text{method\_match}}$ are almost equally important. All other scores play a relatively minor role. 

\section{Discussion}
In this section, we discuss the results of our experiments, the practicality of Broccoli and provide recommendations for Researchers and Practitioners. 
\label{sec:discussion}

\subsection{Search engines for bug localization}

In summary, we accept \textbf{HT1}, since Broccoli did perform significantly better than the other approaches in 3 of 4 experiments and in the reminding case it is in the leading group. Thus, the additional search engine reveals information from the bug report and the source code, which is not provided by the other bug localization techniques. A general search engine can be used for locating potential bug locations. Using the search engine without common bug localization techniques like stack trace detection, the performance would be lower compared to the state-of-the-art techniques. However, the \gls{MRR} can only be increased in one of two experiments. Considering the intuition behind the metrics, the search engine is able to provide the software developer more relevant matches in the top-10 hits, while the first match is at least at the same position compared to the other approaches. Overall, adding a search engine can only increase the result of the bug localization.    

We validated our findings by calculating the feature importance in both data sets. The search engine component that uses the content of our source code files is the most relevant feature. This underlines that the search engine is able to find bug locations and furthermore it is the most relevant component in the Random Forest. Thus, the search engine can locate bugs more precisely than previous approaches. However, the feature importance of other components can be reduced by the search engine component, if the search engine tries to find the same amount of information. For example, the search engine finds class, method and file names better than the regular expression. This indicates that the search engine is a better feature for than the regular expression for class, method and file name detection.    

A potential reason for the different result in the first and second phase of our experiment can be a disparate data collection technique. Lee et al. ~\cite{lee} emphasized the importance of the correct source code selection and the quality of the data set for benchmarking. In the second data set, we used manually validated data of the SmartSHARK data set, which raised the quality of the data. Other studies, that underline that wrongly classified bug reports or bloated ground truths do not statistically significantly impact the bug localization results  ~\cite{kochhar2014potential} ~\cite{mills2020relationship}. Nevertheless, both data sets contain different projects. The way how bug reports are managed could also influence the performance of the bug localization. For example, if the bug reports in one project contain many stack traces or class names that helps to find the bug location.

\begin{mdframed}
\textbf{RQ1 and HT1:} We accept HT1 that a search engine is able to improve mean performance of the bug localization. However, the improvement is only for significant for \gls{MAP}. For \gls{MRR}, there is a comparable performance.
\end{mdframed}

\subsection{Impact of Time-awareness}

Considering the data and Figure \ref{fig:experiment3}, it is noticeable, that for some projects, in which the Time-Aware matching outperforms the major version matching strategy, the Time-Aware matching values are substantially greater. However, if the release approach matching strategy outperforms the Time-Aware matching, the differences between the two values are usually rather small. A possible explanation could be the number of bug reports that are corrected and how the fixed files of the bug report are divided among the different source code versions. 

In summary, the statistical analysis and the results of the two experiments indicate that we can accept \textbf{HT2}, because we can reduce the number of non-detectable source code files of bug reports. The results indicate that the usage of major releases is a lower bound to the performance of the bug localization approach in an application. Therefore, the strategy of using major releases is a good indicator for the practical performance of the bug localization approach. The exact number differs for each project and depends for example on the number of release versions and bug distribution. It is noticeable that factors like massive refactoring or package renaming influence the benchmarking dramatically. The main disadvantage of the Time-Aware strategy is the runtime. In our experiment on the SmartSHARK data, the runtime increased by factor 3-4. However, this is only a problem for researchers, as there would be no need for similar considerations in an actual deployment of a bug localization model in practice. 

\begin{mdframed}
\textbf{RQ2 and HT2:}  We can accept \textbf{HT2}, since there is a a significant increase in the performance metrics \gls{MAP} and \gls{MRR}.
\end{mdframed}

\subsection{Recommendations for Researchers}

Our results show that there are disadvantages in the commonly used data for bug localization benchmarking. However other studies emphasize that wrongly classified bug reports or bloated ground truths do not impact the bug localization results in a statistically significant way~\cite{kochhar2014potential} ~\cite{mills2020relationship}.
We recommend that potential future work should investigate: 

\begin{itemize}
\item \textbf{Number of features:} Figure \ref{fig:featureImportance} shows that the version history component is just an overfitting for a project rather than a bug localization technique that should be used in general. A future bug localization approach could try to reduce the number of components in the Random Forest to only use the most relevant components. This follows the principle of simplicity. 

\item \textbf{Relevance of \gls{MAP} and \gls{MRR}:} The literature did not specify, which metric is more relevant for the quality of the result for a software developer to locate bugs. A user study would be helpful to investigate the relevance of the metrics and help to rank future approaches in a more practical orientated metric. 

\item \textbf{Extend the time-aware strategy:} Due to computing complexity, we were not able to use all revisions of the software project as a potential search space. Thus, the Time-Aware matching strategy was not able to find all files mentioned in the bug reports. However, to use a more realistic approach for benchmarking, we recommend to use any revision prior to the bug report date as a potential bug location. Extending this idea, it would be helpful to track file name changes or at least provide unified identifiers for each file in a project.  

\end{itemize}

\subsection{Recommendations for Practitioners}

Generally, the results of our experiment show that none of the algorithms can be selected without additional knowledge of the project in order to efficiently locate a bug. Each of the algorithm depends on the availability of historical data and how a bug report is typically structured. In detail, the result of our second phase shows that the quality of a bug report is relevant for the software developer as well as for the bug localization approach. 
We recommend that practitioners will consider the following ideas in their operationalizations of bug localization: 

\begin{itemize}

\item \textbf{Bug report quality:} As indicated by our results, the general requirement, a good structure and as much as possible information about the bug in the report is directly correlated to the quality of the bug localization result. For example stack traces or method names improved the result of the bug localization.

\item \textbf{Project specific adjustment:} The scoring part of the bug localization approach should be trained directly on the project data set. This helps to adjust the bug localization algorithm for the project by selecting relevant components. Additionally, we recommend to select the relevant branches beforehand of the software project to reduce runtime issues by simultaneously providing a large, searchable corpus.

\end{itemize}

\section{Threads to validity}
In this section, we evaluate threats to validity of our experiment. We separate between the construct, the internal and the external validity. 
\label{sec:threats}

\subsection{Construct Validity}

The main threat to construct validity lies in the implementation of the other algorithms and the metrics we used to compare the results of our experiments. We tried to reduce this threat by using a replication kit and the implementation from the benchmark study Bench4BL ~\cite{lee}. All implementations of the approaches were taken from the study, except Blizzard, since they provided their own replication kit~\cite{masud_rahman_2018_1297907}. We carefully reviewed the replication kits and only refactored the source code to fit in our benchmark tool, without changing the functionality of the algorithms. Moreover, we used two relevant information retrieval metrics, which were already used by many past bug localization approaches, for example ~\cite{lee} ~\cite{rahman2018improving}  ~\cite{wen2016locus}. Furthermore, the benchmark Bench4BL ~\cite{lee} uses these metrics to compare the approaches. Hence, we believe to have reduced the threat to construct validity to minimum.

\subsection{Internal Validity}

The main threat to internal validity lies in the power of the statistical procedures used to measure the difference between the approaches. However, this would not affected the general results of our work, rather the ranking of the approaches.
Moreover, there may be some noise in the data due to wrong links between bug fixing commits and bug issues. However, this was manually validated for the SmartSHARK data and the authors of Bench4BL designed their data collection to err on the side of caution and avoid wrong links. Similarly, not all issues reported as bug are actually bugs~\cite{Herzig2013}. While the types of issues in the SmartSHARK data were validated, no such manual validation was conducted for the Bench4BL data. Kochar\etal~\cite{kochhar2014potential} showed that this may have a small but significant impact on our results for the Bench4BL data.

\subsection{External Validity}

The main threat to external validity lies in the two data sets that we used in our experiments. The data sets contains only open source projects and are restricted to Java programming language. To reduce the influence, we selected a huge number of open source projects. Overall, we executed our benchmark tool on 82 open source projects, which is currently the largest data set in the area of bug localization. Therefore, the empirical results should not be too specific. Moreover, Ma et al. ~\cite{ma2013commercial} outlined that many issues in open source software projects are reported by commercial developers. Therefore, the performance of the bug localization approaches should be similar on closed source projects. The programming language restriction

\section{Conclusion} 
\label{chap:ende}

Our contribution consists of three parts: First, we presented Broccoli, a new bug localization approach using a off-the-shelf search engine to improve overall performance. Second, we performed benchmarking of eight different bug localization approaches using data from 82 open source projects. Third, we outlined a new strategy for correct source code selection, called Time-Aware matching, which allows a supplementary real-world benchmarking and validated the strategy with a data set of 38 projects.

Our bug localization approach, Broccoli, uses well-known techniques of information retrieval to locate buggy files. These include, regular expressions for stack trace retrieval and natural language preprocessing like stemming or lemmatization. In addition, Broccoli uses advanced techniques like building an abstract syntax tree to extract information. On top of that, Broccoli uses a search engine to locate the source code files most likely to contain the bug. Finally, we trained our approach with a Random Forest algorithm using unrelated software projects to get a joint decision based on the scores of the different information retrieval components.

To evaluate our approach we performed two experiments with state-of-the art bug localization algorithms and compare their performance against Broccoli using the common metrics \gls{MAP} and \gls{MRR}. We performed a statistical analysis of the results to compare the approaches. 
For the first phase of the experiment, we used the proposed data set of Lee et al. ~\cite{lee}, who already compared the performance of six bug localization algorithms on 51 open source software projects. We extended the study by adding Blizzard and Broccoli to the comparison. The result reveals that none of the approaches works best across all  projects. However, the data indicates that Broccoli outperforms other approaches when the history of the software project is sufficient. Furthermore, the Random Forest of Broccoli can be adjusted to the software project statistics to achieve a higher performance, which makes this algorithm even more relevant in a practical setting. On average, Broccoli accomplished a higher \gls{MRR} and \gls{MAP} value than state-of-the-art approaches. However, the indicated difference is not significant.

In the second phase of the experiment, we collected data using the SmartSHARK platform of 31 additional open source software projects. Then, we compared the eight bug localization approaches regarding their \gls{MAP} and \gls{MRR} value on the data using a Random Forest model trained from phase 1. Similar to our first experiment, the report of the statistical analysis did not propose a single approach to be the statistically better than the others. However, the second experiment reveals different groups between the approaches, which are significantly different to each other. The leading group consists of Broccoli, Blizzard and BRTracer+ for the \gls{MAP} metric and Broccoli and BRTracer+ for the \gls{MRR} metric. The results show that Broccoli dominates the results for the \gls{MAP} value, since the approach outperforms the other approaches in 16 of 31 projects. Our results indicate a significant difference between Broccoli and all other approaches regarding the \gls{MAP} metric. The performance of BRTracer+ regarding the \gls{MRR} metric is superior to Broccoli.

The Time-Aware matching strategy helps selecting a more accurate source code version regarding a bug report. Lee et al. ~\cite{lee} already outlined the relevance of the correct selection regarding the performance of bug localization. With our experiment we have shown that the Time-Aware matching is able to influence the \gls{MAP} value and the \gls{MRR} value of Broccoli significantly. Generally, the experiment results show that approx. 23\% of the source code files can not be detected by using the common strategy, since the source files are not included in the search space. Moreover, 11\% of the bug reports can not be predicted, since none of the fixed files are included in the source code of the corresponding release version. Time-Aware matching was able to decrease the non-detectable files to 14\% and the non-detectable bug reports to 0.3\%. This increase comes with the price of extended runtime to calculate the appropriate data. Depending on the data set, the runtime can increase three to four times.

\section{Outlook and future work}
Broccoli can be extended, for example, by improving the scoring step. Our experiment results indicate that a second machine learning model could improve the localization even more. This model should be able to assess project specific properties and adjust the scoring to the actual project. The second model and the general Random Forest model are then combined by a new function. Furthermore, the search engine can be improved by using word embeddings that are trained with the programming language. The principle of query reformulation can also be used to improve the result of the search engine. Mills et al. ~\cite{mills2020relationship} already examined the relevance of the query reformulation.

Additionally, we would like to increase the granularity of Broccoli. Thus, we want to provide the software developer with information about the code range where the bug is located. In order to implement this, additional experiments regarding method and variable names as well as Java documentation are necessary. Moreover, the quality of the ground truth in the data set must be extended to the code line level ~\cite{mills2020relationship}. Hence, future work should provide a large data set of manually validated bug fixing changes.

Furthermore, the Broccoli approach can be extended to other programming languages, since we do not use special properties of the Java language in our approach. A possible language should be object-oriented and have a package or folder structure. It would be helpful, if the language reports crash reports or stack traces. These criteria are met by most of the popular programming languages. Future work could then also reveal differences between programming languages regarding the relevance of the components.

Regarding the benchmarking of bug localization approaches, different strategies to create a data set that reflects the real application should be analyzed. We already proposed the Time-Aware strategy to find better source code versions that correspond to the bug report. However, not all modified files of the bug reports are included in one of the source code versions. Therefore, future work can extend the search space of source code files to a finer granularity than the set of release versions. Additionally, the source code selection strategy needs further improvements regarding the run time. In the future, it is necessary to evaluate and analyse a more sophisticated strategy, which does not depend on release version but rather on the actual commit history. Moreover, we would like to extend the experiments of the Time-Aware strategy to the data set of Lee et al. ~\cite{lee}.

Furthermore, we like to reduce the threats to external validity further by extending the benchmarking to more bug reports and other software projects. Additionally, we would like to reduce the threats to construct validity by performing a user study on how well a bug localization approach helps to locate the bug in a real-world application.

\bibliographystyle{unsrt}
\bibliography{main}

\appendix

\section{Detailed Results}
\label{sec:details}

\begin{table}[]
\centering
\resizebox{\textwidth}{!}{%
\begin{tabular}{llrrrrrrrr}
\toprule
   Group &        Project &  BugLocator &  BRTracer+ &   BLUiR+ &  AmaLgam+ &    BLIA &   Locus &  Blizzard &  Broccoli \\
\midrule
   Apache &          CAMEL &      0.3585 &    \cellcolor[HTML]{e7e7e7} 0.3960 &  0.3487 &   0.3525 &  0.3479 &  0.4277 &  0.253300 &  0.395668 \\ 
   Apache &          HBASE &      0.3626 &    0.4004 &  0.3533 &   0.3531 &  0.3823 &  0.4322 &  0.282126 &  \cellcolor[HTML]{e7e7e7} 0.430027 \\
   Apache &           HIVE &      0.3048 &    0.3517 &  0.3177 &   0.3181 &  0.2888 &  \cellcolor[HTML]{e7e7e7}0.3546 &  0.219608 &  0.323904 \\
   Commons &          CODEC &      0.6390 &    0.6237 &  0.7488 &   0.7488 &  0.6734 &  \cellcolor[HTML]{e7e7e7} 0.7516 &  0.427676 &  0.710430 \\
   Commons &    COLLECTIONS &      0.4443 &    0.5236 &  0.4057 &   0.4057 &  0.5692 &  \cellcolor[HTML]{e7e7e7} 0.6602 &  0.346136 &  0.608212 \\
   Commons &       COMPRESS &      0.5414 &    0.5484 &  0.4839 &   0.4794 &  0.5791 &  0.6109 &  0.369854 &  \cellcolor[HTML]{e7e7e7} 0.650715 \\
   Commons &  CONFIGURATION &      0.5174 &    0.5103 &  0.5063 &   0.5143 &  0.5750 &  0.6095 &  0.330659 &  \cellcolor[HTML]{e7e7e7} 0.635993 \\
   Commons &         CRYPTO &      0.1628 &    0.1708 &  0.1982 &   0.1982 &  0.1588 &  0.1118 &  0.156250 &  \cellcolor[HTML]{e7e7e7} 0.277239 \\
   Commons &            CSV &      0.6017 &    0.5813 &  0.6052 &   0.6290 &  0.3750 &  0.6554 &  0.440476 &  \cellcolor[HTML]{e7e7e7} 0.759524 \\  
   Commons &             IO &      0.7110 &    0.7152 &  0.6694 &   0.6693 &  0.7629 &  \cellcolor[HTML]{e7e7e7} 0.7943 &  0.405193 &  0.780702 \\ 
   Commons &           LANG &      0.5576 &    0.5648 &  0.5674 &   0.5714 &  0.5988 &  0.6737 &  0.394735 &  \cellcolor[HTML]{e7e7e7} 0.724163 \\
   Commons &           MATH &      0.4227 &    0.4459 &  0.4834 &   0.4965 &  0.4696 &  0.4756 &  0.342118 &  \cellcolor[HTML]{e7e7e7} 0.586698 \\
   Commons &         WEAVER &      0.6212 &    0.6331 &  0.6637 &  \cellcolor[HTML]{e7e7e7}  0.6637 &  0.5694 &  0.3581 &  0.312500 &  0.470810 \\
   JBoss &         ENTESB &    \cellcolor[HTML]{e7e7e7}  0.0999 &    0.0875 &  0.0949 &   0.1546 &  0.0539 &  0.1178 &  0.033466 &  0.066006 \\
   JBoss &         JBMETA &      0.2272 &    0.2332 &  0.1685 &   0.1706 &  0.2078 &  \cellcolor[HTML]{e7e7e7} 0.2717 &  0.128016 &  0.242185 \\
   Wildfly &            ELY &      0.0625 &    0.0672 &  0.1098 &   0.1098 &  0.0939 &  0.1270 &  0.036936 &  \cellcolor[HTML]{e7e7e7} 0.557435 \\
   Wildfly &          SWARM &      0.2857 &    0.2967 &  0.2778 &   0.2778 &  0.2522 &  0.2849 &  0.152985 & \cellcolor[HTML]{e7e7e7} 0.486699 \\
   Wildfly &          WFARQ &      0.5000 &    0.3333 & \cellcolor[HTML]{e7e7e7} 1.0000 &  \cellcolor[HTML]{e7e7e7} 1.0000 &  0.1111 &  0.5000 &  0.290096 &  0.598045 \\
   Wildfly &         WFCORE &      0.3225 &    0.3353 &  0.2693 &   0.2706 &  0.2826 &  0.3633 &  0.227879 &  \cellcolor[HTML]{e7e7e7} 0.496164 \\
   Wildfly &           WFLY &      0.3036 &    0.3337 &  0.2698 &   0.2706 &  0.2624 &  0.3542 &  0.302594 & \cellcolor[HTML]{e7e7e7} 0.504423 \\
   Wildfly &           WFMP &      0.4535 &    0.2675 & \cellcolor[HTML]{e7e7e7} 0.8485 &   0.8485 &  0.5000 &  0.6226 &  0.262787 &  0.130890 \\
   Spring &           AMQP &      0.4136 &    0.4369 &  0.4275 &   0.4306 &  0.4532 &  0.4864 &  0.279517 & \cellcolor[HTML]{e7e7e7} 0.579131 \\
   Spring &        ANDROID &      0.4016 &    0.3238 &  0.3421 &   0.3421 &  0.3195 & \cellcolor[HTML]{e7e7e7} 0.5405 &  0.279532 &  0.491204 \\
   Spring &          BATCH &      0.3535 &    0.3685 &  0.3738 &   0.3808 &  0.3845 &  0.4234 &  0.241711 & \cellcolor[HTML]{e7e7e7} 0.672407 \\
   Spring &       BATCHADM &      0.3825 &    0.4095 &  0.3898 &   0.3898 & \cellcolor[HTML]{e7e7e7} 0.4602 &  0.4513 &  0.068583 &  0.458225 \\
   Spring &       DATACMNS &      0.3649 &    0.3766 &  0.3631 &   0.3630 &  0.4319 &  0.4692 &  0.290383 & \cellcolor[HTML]{e7e7e7} 0.542476 \\
   Spring &      DATAGRAPH &      0.0396 &    0.0494 &  0.0352 &   0.0352 &  0.0465 &  0.0667 &  0.245129 &  \cellcolor[HTML]{e7e7e7} 0.566223 \\
   Spring &        DATAJPA &      0.4022 &    0.4150 &  0.4221 &   0.4236 & \cellcolor[HTML]{e7e7e7} 0.4816 &  0.4106 &  0.337211 &  0.297177 \\
   Spring &      DATAMONGO &      0.4106 &    0.4412 &  0.4111 &   0.4111 &  0.4724 & \cellcolor[HTML]{e7e7e7} 0.4747 &  0.250484 &  0.395668 \\
   Spring &      DATAREDIS &      0.5255 &    0.5478 &  0.5523 &   0.5559 &  \cellcolor[HTML]{e7e7e7} 0.5571 &  0.4905 &  0.280649 &  0.535546 \\
   Spring &       DATAREST &      0.3377 &    0.3600 &  0.3297 &   0.3335 &  0.3511 &  0.3094 &  0.329149 & \cellcolor[HTML]{e7e7e7} 0.468639 \\
   Spring &           LDAP &      0.4697 &    0.5393 &  0.5017 &   0.5017 &  0.4948 &  0.3821 &  0.211100 &  \cellcolor[HTML]{e7e7e7} 0.556067 \\
   Spring &         MOBILE &      0.6697 &    0.7057 & \cellcolor[HTML]{e7e7e7} 0.9264 &  \cellcolor[HTML]{e7e7e7} 0.9264 &  0.7316 &  0.4519 &  0.300328 &  0.552407 \\
   Spring &            ROO &      0.4111 &    0.4436 &  0.3874 &   0.3970 &  0.4168 &  0.4019 &  0.172328 & \cellcolor[HTML]{e7e7e7} 0.617940 \\
   Spring &            SEC &      0.4606 &    0.4822 &  0.4421 &   0.4480 &  0.5188 &  0.5126 &  0.302329 & \cellcolor[HTML]{e7e7e7} 0.757648 \\
   Spring &       SECOAUTH &      0.3300 &    0.3233 &  0.2975 &   0.3100 &  0.3153 &  0.3244 &  0.377671 & \cellcolor[HTML]{e7e7e7} 0.712500 \\
   Spring &            SGF &      0.3876 &    0.3717 &  0.3899 &   0.3898 &  0.3808 &  0.3542 &  0.209392 & \cellcolor[HTML]{e7e7e7} 0.636332 \\
   Spring &           SHDP &      0.4127 &    0.4433 &  0.3500 &   0.3500 &  0.4341 & \cellcolor[HTML]{e7e7e7} 0.4501 &  0.302222 &  0.296042 \\
   Spring &            SHL &      0.2802 &    0.2363 &  0.3238 &   0.3239 &  0.3160 &  0.2100 &  0.292963 & \cellcolor[HTML]{e7e7e7} 0.429801 \\
   Spring &         SOCIAL &      0.6108 &    0.6237 &  0.4423 &   0.4342 &  0.5916 & \cellcolor[HTML]{e7e7e7} 0.6633 &  0.314815 &  0.418222 \\
   Spring &       SOCIALFB &      0.6040 &   \cellcolor[HTML]{e7e7e7} 0.6194 &  0.5003 &   0.4804 &  0.4610 &  0.5418 &  0.362500 &  0.195652 \\
   Spring &       SOCIALLI &   \cellcolor[HTML]{e7e7e7}   0.4711 &    0.6384 &  0.3929 &   0.4504 &  0.2990 &  0.4081 &  0.244428 &  0.298814 \\
   Spring &       SOCIALTW &      0.4917 &    0.4941 &  0.5789 & \cellcolor[HTML]{e7e7e7}  0.6622 &  0.3115 &  0.5456 &  0.277495 &  0.250000 \\
   Spring &            SPR &      0.3372 &   \cellcolor[HTML]{e7e7e7} 0.3708 &  0.2380 &   0.2483 &  0.2985 &  0.3573 &  0.215859 &  0.302871 \\
   Spring &            SWF &      0.3708 &    0.3912 &  0.3706 &   0.3739 & \cellcolor[HTML]{e7e7e7} 0.3916 &  0.4415 &  0.112077 &  0.260377 \\
   Spring &            SWS &      0.3515 &    0.3504 &  0.3391 &   0.3391 &  0.3579 & \cellcolor[HTML]{e7e7e7} 0.3661 &  0.128083 &  0.297222 \\
   Old &        AspectJ &      0.2148 &    0.2250 &  0.1428 &   0.2023 & \cellcolor[HTML]{e7e7e7} 0.2488 &  0.1605 &  0.000000 &  0.069830 \\
   Old &            JDT &      0.1697 &    0.2774 &  0.2491 &   0.2491 &  0.1742 &  0.3594 &  0.182133 & \cellcolor[HTML]{e7e7e7} 0.313034 \\
   Old &            PDE &      0.3757 &  \cellcolor[HTML]{e7e7e7}  0.3884 &  0.3330 &   0.3334 &  0.3376 &  0.3711 &  0.152995 &  0.281645 \\
   Old &            SWT &      0.4458 &    0.5257 & \cellcolor[HTML]{e7e7e7} 0.5528 &   0.5528 &  0.4639 &  0.3303 &  0.416667 &  0.323040 \\
   Old &          ZXing &      0.3306 &    0.3937 & \cellcolor[HTML]{e7e7e7} 0.4598 &   0.4596 &  0.4984 &  0.4358 &  0.339949 &  0.393447 \\
   
  \midrule
     Statistic & mean value  & 0.399$\pm$0.084 &  0.412$\pm$0.087 & 0.417$\pm$0.110 & 0.424$\pm$0.109 & 0.394$\pm$0.092 & 0.426$\pm$0.091 & 0.259$\pm$0.058 & 0.459$\pm$0.103  \\
     {} & standard deviation & 0.149 & 0.154 & 0.195 &  0.194 & 0.163 & 0.161 & 0.103 & 0.182 \\
\bottomrule
\end{tabular}%
}
\caption{\gls{MAP} values of the different bug localization approaches using the Bench4BL data set ~\cite{lee}}
\label{tab:map_experiment1}
\end{table}

\begin{table}[]
\centering
\resizebox{\textwidth}{!}{%
\begin{tabular}{llrrrrrrrr}
\toprule
   Group &        Project &  BugLocator &  BRTracer+ &   BLUiR &  AmaLgam+ &    BLIA &   Locus &  Blizzard &  Broccoli \\
\midrule
  Apache &          CAMEL &      0.4658 &    0.5280 &  0.4511 &   0.4550 &  0.4566 & \cellcolor[HTML]{e7e7e7} 0.5524 &  0.250616 &  0.489204 \\
  Apache &          HBASE &      0.4433 &    0.4918 &  0.4284 &   0.4280 &  0.4576 &  0.5025 &  0.278388 & \cellcolor[HTML]{e7e7e7} 0.528472 \\
  Apache &           HIVE &      0.3848 &   \cellcolor[HTML]{e7e7e7}  0.4548 &  0.4098 &   0.4095 &  0.3627 &  0.4497 &  0.216124 &  0.427103 \\
 Commons &          CODEC &    \cellcolor[HTML]{e7e7e7}  0.8230 &    0.7554 &  0.8016 &   0.8016 &  0.8241 &  0.7935 &  0.425813 &  0.763772 \\
 Commons &    COLLECTIONS &      0.6296 &    0.6933 &  0.4581 &   0.4581 & \cellcolor[HTML]{e7e7e7} 0.7252 &  0.7208 &  0.339956 &  0.647919 \\
 Commons &       COMPRESS &      0.7165 &    0.7561 &  0.6139 &   0.6097 & \cellcolor[HTML]{e7e7e7} 0.7871 &  0.7693 &  0.369249 &  0.733370  \\
 Commons &  CONFIGURATION &    \cellcolor[HTML]{e7e7e7}  0.6846 &    0.7088 &  0.5433 &   0.5498 &  0.7558 &  0.5911 &  0.332878 &  0.674996  \\
 Commons &         CRYPTO &      0.2639 &    0.3021 & \cellcolor[HTML]{e7e7e7} 0.3911 &  \cellcolor[HTML]{e7e7e7} 0.3911 &  0.2951 &  0.1059 &  0.138889 &  0.328125  \\
 Commons &            CSV &      0.6434 &    0.6032 &  0.5952 &   0.5952 &  0.5357 &  0.6596 &  0.428571 & \cellcolor[HTML]{e7e7e7} 0.780952 \\
 Commons &             IO &      0.7917 &  \cellcolor[HTML]{e7e7e7}  0.8316 &  0.6680 &   0.6678 &  0.8274 &  0.7550 &  0.401867 &  0.808956 \\
 Commons &           LANG &      0.6154 &    0.6225 &  0.5601 &   0.5580 &  0.6898 &  0.6050 &  0.389248 &  \cellcolor[HTML]{e7e7e7} 0.736603 \\
 Commons &           MATH &      0.4851 &    0.5416 &  0.5289 &   0.5410 &  0.5558 &  0.4508 &  0.335474 & \cellcolor[HTML]{e7e7e7} 0.647067  \\
 Commons &         WEAVER &      0.4167 &    0.3750 & \cellcolor[HTML]{e7e7e7} 0.6667 &  \cellcolor[HTML]{e7e7e7} 0.6667 & \cellcolor[HTML]{e7e7e7} 0.6667 &  0.2917 &  0.291667 &  0.600000  \\
   JBoss &         ENTESB &      0.0515 &    0.0309 &  0.0548 &   0.1256 &  0.0301 & \cellcolor[HTML]{e7e7e7} 0.0748 &  0.035287 &  0.071880 \\
   JBoss &         JBMETA &     \cellcolor[HTML]{e7e7e7} 0.3508 &   \cellcolor[HTML]{e7e7e7} 0.3570 &  0.2500 &   0.2510 &  0.3153 &  0.3530 &  0.121766 &  0.255896  \\
 Wildfly &            ELY &      0.1540 &    0.1587 &  0.1962 &   0.1962 &  0.1667 &  0.1733 &  0.038030 & \cellcolor[HTML]{e7e7e7} 0.643282 \\
 Wildfly &          SWARM &      0.2996 &    0.3741 &  0.3082 &   0.3082 &  0.3189 &  0.3590 &  0.157977 & \cellcolor[HTML]{e7e7e7} 0.677398 \\
 Wildfly &          WFARQ &      0.5000 &    0.3333 & \cellcolor[HTML]{e7e7e7} 1.0000 &  \cellcolor[HTML]{e7e7e7} 1.0000 &  0.1111 &  0.5000 &  0.276234 &  0.698485  \\
 Wildfly &         WFCORE &      0.3888 &    0.4107 &  0.3176 &   0.3187 &  0.3368 &  0.4256 &  0.228333 & \cellcolor[HTML]{e7e7e7} 0.575930  \\
 Wildfly &           WFLY &      0.3483 &    0.3877 &  0.3303 &   0.3360 &  0.3040 &  0.4080 &  0.290186 &  \cellcolor[HTML]{e7e7e7} 0.564736  \\
 Wildfly &           WFMP &      0.5833 &    0.2398 & \cellcolor[HTML]{e7e7e7} 1.0000 &  \cellcolor[HTML]{e7e7e7} 1.0000 &  0.6667 &  0.7778 &  0.240741 &  0.166153  \\
  Spring &           AMQP &      0.5751 &    0.6005 &  0.5549 &   0.5564 &  0.6064 &  0.5745 &  0.276119 & \cellcolor[HTML]{e7e7e7} 0.615128  \\
  Spring &        ANDROID &      0.4612 &    0.3462 &  0.2883 &   0.2883 &  0.4221 &  0.4775 &  0.279532 & \cellcolor[HTML]{e7e7e7} 0.533306 \\
  Spring &          BATCH &      0.4837 &    0.5188 &  0.4572 &   0.4610 &  0.5159 &  0.5286 &  0.242466 & \cellcolor[HTML]{e7e7e7} 0.705288  \\
  Spring &       BATCHADM &      0.4589 &    0.5081 &  0.4170 &   0.4170 &  0.5614 & \cellcolor[HTML]{e7e7e7} 0.5913 &  0.067273 &  0.504142  \\
  Spring &       DATACMNS &      0.4937 &    0.5287 &  0.4367 &   0.4367 &  0.5692 &  0.5396 &  0.291230 & \cellcolor[HTML]{e7e7e7} 0.611770  \\
  Spring &      DATAGRAPH &      0.0503 &    0.0853 &  0.0429 &   0.0429 &  0.0681 &  0.1198 &  0.246244 & \cellcolor[HTML]{e7e7e7} 0.608745  \\
  Spring &        DATAJPA &      0.5216 &  0.5644 &  0.5080 &   0.5102 &  \cellcolor[HTML]{e7e7e7} 0.6331 &  0.4890 &  0.333425 &  0.362282  \\
  Spring &      DATAMONGO &      0.4965 &    0.5379 &  0.4553 &   0.4553 &  \cellcolor[HTML]{e7e7e7} 0.5545 &  0.5341 &  0.246916 &  0.481404  \\
  Spring &      DATAREDIS &      0.7356 &   \cellcolor[HTML]{e7e7e7} 0.7873 &  0.7503 &   0.7539 &  0.7439 &  0.6823 &  0.275696 &  0.589753 \\
  Spring &       DATAREST &      0.5184 &    0.5534 &  0.4627 &   0.4596 &  0.5562 &  0.4521 &  0.329149 & \cellcolor[HTML]{e7e7e7} 0.605260 \\
  Spring &           LDAP &      0.5782 &    0.6780 &  0.5645 &   0.5644 &  0.6149 &  0.4520 &  0.208391 & \cellcolor[HTML]{e7e7e7} 0.682101  \\
  Spring &         MOBILE &      0.5411 &    0.6369 & \cellcolor[HTML]{e7e7e7} 0.6727 &  \cellcolor[HTML]{e7e7e7} 0.6727 &  0.5379 &  0.4218 &  0.300562 &  0.572897  \\
  Spring &            ROO &      0.5049 &    0.5530 &  0.4787 &   0.4882 &  0.5103 &  0.4870 &  0.171809 & \cellcolor[HTML]{e7e7e7} 0.645716  \\
  Spring &            SEC &      0.5727 &    0.6049 &  0.4900 &   0.4978 &  0.6333 &  0.5793 &  0.300756 & \cellcolor[HTML]{e7e7e7} 0.777778  \\
  Spring &       SECOAUTH &      0.4115 &    0.4078 &  0.3495 &   0.3543 &  0.4292 &  0.4362 &  0.377642 & \cellcolor[HTML]{e7e7e7} 0.712500  \\
  Spring &            SGF &      0.5794 &    0.5642 &  0.5283 &   0.5282 &  0.5571 &  0.5358 &  0.201058 & \cellcolor[HTML]{e7e7e7} 0.658554  \\
  Spring &           SHDP &      0.5264 &    \cellcolor[HTML]{e7e7e7} 0.5570 &  0.4146 &   0.4146 &  0.5131 &  0.5246 &  0.296667 &  0.397107  \\
  Spring &            SHL &      0.4493 &    0.4183 &  0.4756 &   0.4757 & \cellcolor[HTML]{e7e7e7} 0.5152 &  0.2800 &  0.292963 &  0.481131  \\
  Spring &         SOCIAL &      0.6608 &    0.6733 &  0.5092 &   0.4764 &  0.6667 & \cellcolor[HTML]{e7e7e7} 0.7030 &  0.314815 &  0.503739 \\
  Spring &       SOCIALFB &   \cellcolor[HTML]{e7e7e7}   0.6744 &    0.7400 &  0.5274 &   0.5330 &  0.4743 &  0.5374 &  0.362500 &  0.202899 \\
  Spring &       SOCIALLI &      0.5000 &    0.5833 &  0.2708 &   0.2917 &  0.2375 & \cellcolor[HTML]{e7e7e7} 0.6250 &  0.240285 &  0.381590 \\
  Spring &       SOCIALTW &      0.5698 &    0.5804 &  0.5833 &  \cellcolor[HTML]{e7e7e7} 0.7083 &  0.3854 &  0.5833 &  0.275533 &  0.250000 \\
  Spring &            SPR &      0.4230 &   \cellcolor[HTML]{e7e7e7} 0.4530 &  0.3425 &   0.3504 &  0.3760 &  0.4415 &  0.209180 &  0.382005 \\
  Spring &            SWF &      0.4576 &    0.4932 &  0.4565 &   0.4586 &  0.4854 & \cellcolor[HTML]{e7e7e7} 0.5153 &  0.110628 &  0.338143 \\
  Spring &            SWS &      0.4641 &    0.4849 &  0.4458 &   0.4458 &  \cellcolor[HTML]{e7e7e7} 0.5011 &  0.4700 &  0.128381 &  0.283333 \\
     Old &        AspectJ &      0.3857 &    0.4161 &  0.2587 &   0.3382 & \cellcolor[HTML]{e7e7e7} 0.4155 &  0.2186 &  0.000000 &  0.151042 \\
     Old &            JDT &      0.2699 &    0.3832 &  0.3382 &   0.3382 &  0.2310 &  0.4542 &  0.183596 & \cellcolor[HTML]{e7e7e7} 0.472323 \\
     Old &            PDE &      0.4971 &    \cellcolor[HTML]{e7e7e7} 0.5382 &  0.4304 &   0.4308 &  0.4714 &  0.5236 &  0.153110 &  0.378770 \\
     Old &            SWT &      0.5016 &    0.5967 & \cellcolor[HTML]{e7e7e7} 0.6286 &   0.6284 &  0.5666 &  0.3812 &  0.388889 &  0.487189 \\
     Old &          ZXing &      0.3837 &    0.4219 &  0.5626 &   0.5626 & \cellcolor[HTML]{e7e7e7} 0.5752 &  0.5274 &  0.353940 &  0.464660 \\
     
     \midrule
     Statistic & mean value  & 0.486$\pm$0.090 &  0.505$\pm$0.095 &  0.476$\pm$0.105 & 0.482$\pm$0.103 & 0.492$\pm$0.105 & 0.490$\pm$0.093 & 0.256$\pm$0.057 & 0.523$\pm$0.102  \\
     {} & standard deviation & 0.159 & 0.169 & 0.186  & 0.183 & 0.186 & 0.166 & 0.102 & 0.181  \\
\bottomrule
\end{tabular}
}
\caption{\gls{MRR} values of the different bug localization approaches using the Bench4BL data set ~\cite{lee}}
\label{tab:mrr_experiment1}
\end{table}

\begin{table}[]
\centering
\resizebox{\textwidth}{!}{%
\begin{tabular}{lrrrrrrrr}
\toprule
{}Subject &  Blizzard & BugLocator &  BRTracer+ &     BLUiR+ &   AmaLgam+ &      BLIA &     Locus   &  Broccoli \\
\midrule
ant-ivy  &  0.332016 &    0.390238 &  0.441654 &  0.452082 &  0.132309 &  0.471136 &  0.371438 & \cellcolor[HTML]{e7e7e7} 0.496122 \\
archiva  &  0.283810 &    0.205291 &  0.235622 &  0.309811 &  0.107460 &  0.257067 &  0.179104 & \cellcolor[HTML]{e7e7e7} 0.370623 \\
calcite  & \cellcolor[HTML]{e7e7e7} 0.661462 &    0.362004 &  0.409113 &  0.482388 &  0.089639 &  0.416247 &  0.300452 &  0.561968 \\
cayenne  &  0.318740 &    0.289970 &  0.313624 &  0.320181 &  0.057747 &  0.291917 &  0.182501 & \cellcolor[HTML]{e7e7e7} 0.402604 \\
commons-bcel  &  0.343043 &    0.335842 &  0.323605 &  0.255740 &  0.093842 &  0.340908 &  0.232546 & \cellcolor[HTML]{e7e7e7} 0.425496 \\
commons-beanutils  &  0.638323 &    0.515891 &  0.516848 &  0.549294 &  0.081009 &  0.621000 &  0.260726 & \cellcolor[HTML]{e7e7e7} 0.704911 \\
commons-dbcp &  0.372420 &    0.362946 &  0.407908 &  0.265410 &  0.055933 & \cellcolor[HTML]{e7e7e7} 0.462766 &  0.210462 &  0.399022 \\
commons-digester &  0.465152 &    0.583333 & \cellcolor[HTML]{e7e7e7} 0.638889 &  0.513889 &  0.143894 &  0.353175 &  0.027778 &  0.605556 \\
commons-jcs &  0.241268 &    0.303888 &  0.309193 &  0.198352 &  0.096700 & \cellcolor[HTML]{e7e7e7} 0.362969 &  0.169063 &  0.347422 \\
commons-jexl &  0.299008 &    0.254857 &  0.322224 &  0.309314 &  0.062730 &  0.309210 &  0.241374 & \cellcolor[HTML]{e7e7e7} 0.387787 \\
commons-net &  0.570368 &    0.757696 & \cellcolor[HTML]{e7e7e7} 0.777640 &  0.647473 &  0.201341 &  0.771412 &  0.574736 &  0.774366 \\
commons-scxml &  0.451146 &    0.398328 &  0.500226 &  0.547092 &  0.235204 &  0.595953 &  0.260388 & \cellcolor[HTML]{e7e7e7} 0.666737 \\
commons-validator &  0.537459 &    0.634593 &  0.616823 &  0.690180 &  0.397130 &  0.648970 &  0.455882 & \cellcolor[HTML]{e7e7e7} 0.738481 \\
commons-vfs &  0.347959 &    0.336728 &  0.323547 &  0.335355 &  0.087631 & \cellcolor[HTML]{e7e7e7} 0.382568 &  0.023923 &  0.374173 \\
deltaspike &  0.433986 &    0.436944 &  0.469685 &  0.316029 &  0.092796 &  0.418076 &  0.264562 & \cellcolor[HTML]{e7e7e7} 0.498238 \\
eagle &  0.320228 &    0.248688 &  0.306678 &  0.296099 &  0.093546 &  0.297936 &  0.216504 & \cellcolor[HTML]{e7e7e7} 0.349435 \\
giraph & \cellcolor[HTML]{e7e7e7} 0.561540 &    0.385187 &  0.458477 &  0.219029 &  0.048874 &  0.478083 &  0.270718 &  0.473961 \\
gora &  0.587114 &    0.568033 &  0.639319 &  0.401889 &  0.191640 & \cellcolor[HTML]{e7e7e7} 0.721169 &  0.363479 &  0.610263 \\
jspwiki &  0.095989 &    0.084489 & \cellcolor[HTML]{e7e7e7} 0.126419 &  0.077290 &  0.018410 &  0.117256 &  0.074270 &  0.122252 \\
knox & \cellcolor[HTML]{e7e7e7} 0.593351 &    0.387693 &  0.391028 &  0.442024 &  0.164717 &  0.365978 &  0.305243 &  0.473973 \\
kylin &  0.318255 &    0.367787 &  0.377638 &  0.331271 &  0.096778 &  0.366618 &  0.272751 & \cellcolor[HTML]{e7e7e7} 0.396795 \\
lens &  0.371659 &    0.300424 &  0.364237 &  0.321388 &  0.102869 &  0.340073 &  0.261586 & \cellcolor[HTML]{e7e7e7} 0.416356 \\
mahout &  0.487345 &    0.504056 &  0.492787 &  0.456864 &  0.136894 &  0.505729 &  0.225595 & \cellcolor[HTML]{e7e7e7} 0.556360 \\
manifoldcf &  0.476083 &    0.445142 &  0.538529 & \cellcolor[HTML]{e7e7e7} 0.585691 &  0.240106 &  0.537491 &  0.544708 &  0.547109 \\
nutch &  0.443553 &    0.453102 &  0.441882 &  0.412601 &  0.175980 &  0.471091 &  0.372920 & \cellcolor[HTML]{e7e7e7} 0.483704 \\
opennlp &  0.478717 &    0.467758 &  0.469991 &  0.439160 &  0.148400 &  0.511765 &  0.318779 & \cellcolor[HTML]{e7e7e7} 0.546073 \\
parquet-mr & \cellcolor[HTML]{e7e7e7} 0.508280 &    0.472863 &  0.454522 &  0.442623 &  0.072847 &  0.498007 &  0.208646 &  0.461694 \\
santuario-java &  0.449190 &    0.490680 &  0.529292 &  0.501296 &  0.137956 &  0.512916 &  0.400176 & \cellcolor[HTML]{e7e7e7} 0.622546 \\
systemml & \cellcolor[HTML]{e7e7e7} 0.343453 &    0.223307 &  0.256455 &  0.275636 &  0.088209 &  0.243436 &  0.244510 &  0.340754 \\
tika &  0.520047 &    0.463094 &  0.473150 & \cellcolor[HTML]{e7e7e7} 0.550030 &  0.249915 &  0.482560 &  0.306794 &  0.543742 \\
wss4j &  0.227173 &    0.262678 &  0.266480 &  0.189150 &  0.072923 & \cellcolor[HTML]{e7e7e7} 0.289911 &  0.099424 &  0.268095 \\
\midrule
mean value  & 0.444$\pm$0.217 & 0.388$\pm$0.276 &  0.442$\pm$0.271 & 0.402$\pm$0.251 & 0.097$\pm$0.174 & 0.418$\pm$0.264 &  0.261$\pm$0.273 &  0.474$\pm$0.253 \\
standard deviation & 0.149 & 0.126 & 0.130 & 0.147 & 0.058 &  0.130 & 0.077 & 0.128 \\
\bottomrule
\end{tabular}%
}
\caption{\gls{MAP} values of the different bug localization approaches using the SmartSHARK data set}
\label{tab:map_experiment2}
\end{table}

\begin{table}[]
\centering
\resizebox{\textwidth}{!}{%
\begin{tabular}{lrrrrrrrr}
\toprule
{} Subject &  Blizzard &  BugLocator &  BRTracer+ &     BLUiR+ &   AmaLgam+ &      BLIA &     Locus  &   Broccoli \\
\midrule
ant-ivy  &  0.329682 &    0.507787 &  0.569334 &  0.447456 &  0.165538 & \cellcolor[HTML]{e7e7e7} 0.576928 &  0.464967 &  0.538422 \\
archiva  &  0.282379 &    0.314665 &  0.327339 &  0.301384 &  0.158616 &  0.326369 &  0.269727 & \cellcolor[HTML]{e7e7e7} 0.408797 \\
calcite  &  0.644654 &    0.573255 & \cellcolor[HTML]{e7e7e7} 0.645761 &  0.487828 &  0.146583 &  0.593231 &  0.475130 &  0.615734 \\
cayenne  &  0.314903 &    0.410786 & \cellcolor[HTML]{e7e7e7} 0.439235 &  0.349688 &  0.080382 &  0.378046 &  0.269904 &  0.425274 \\
commons-bcel  &  0.335531 &  \cellcolor[HTML]{e7e7e7}  0.450298 &  0.441657 &  0.251786 &  0.104030 &  0.426032 &  0.279258 &  0.436706 \\
commons-beanutils  &  0.626786 &    0.708285 &  0.727604 &  0.609552 &  0.136817 &  0.709683 &  0.371895 & \cellcolor[HTML]{e7e7e7} 0.765278 \\
commons-dhcp &  0.373523 &    0.448418 &  0.494504 &  0.276410 &  0.071108 & \cellcolor[HTML]{e7e7e7} 0.502436 &  0.286095 &  0.428291 \\
commons-digester &  0.459596 &    0.583333 &  0.666667 &  0.500000 &  0.201038 &  0.472222 &  0.027778 & \cellcolor[HTML]{e7e7e7} 0.700000 \\
commons-jcs &  0.244312 &    0.397585 &  0.405445 &  0.206041 &  0.136943 & \cellcolor[HTML]{e7e7e7} 0.434517 &  0.201739 &  0.375926 \\
commons-jexl &  0.275767 &    0.405581 & \cellcolor[HTML]{e7e7e7} 0.473746 &  0.265904 &  0.114128 &  0.369101 &  0.383866 &  0.436508 \\
commons-net &  0.563316 &    0.830534 & \cellcolor[HTML]{e7e7e7} 0.847132 &  0.626864 &  0.230168 &  0.831587 &  0.611000 &  0.818600 \\
commons-scxml &  0.453175 &    0.525560 &  0.653922 &  0.474157 &  0.366592 & \cellcolor[HTML]{e7e7e7} 0.743697 &  0.378175 &  0.708333 \\
commons-validator &  0.516296 &    0.745397 &  0.693401 &  0.675212 &  0.402892 &  0.729312 &  0.494762 & \cellcolor[HTML]{e7e7e7} 0.792275 \\
commons-vfs &  0.344450 &    0.404811 &  0.381508 &  0.357395 &  0.111512 & \cellcolor[HTML]{e7e7e7} 0.419488 &  0.023923 &  0.388341 \\
deltaspike &  0.427129 &    0.520585 & \cellcolor[HTML]{e7e7e7} 0.574527 &  0.312629 &  0.109748 &  0.470188 &  0.319387 &  0.530922 \\
eagle &  0.325775 &    0.372153 & \cellcolor[HTML]{e7e7e7} 0.456326 &  0.294178 &  0.125067 &  0.385611 &  0.311256 &  0.376595 \\
giraph &  0.552878 &    0.480869 & \cellcolor[HTML]{e7e7e7} 0.565809 &  0.229035 &  0.053924 &  0.538071 &  0.305481 &  0.499654 \\
gora &  0.575475 &    0.657885 &  0.760391 &  0.391488 &  0.246376 & \cellcolor[HTML]{e7e7e7} 0.784314 &  0.416824 &  0.630034 \\
jspwiki &  0.093277 &    0.123673 & \cellcolor[HTML]{e7e7e7} 0.161616 &  0.073212 &  0.026203 &  0.131458 &  0.088165 &  0.133227 \\
knox & \cellcolor[HTML]{e7e7e7} 0.583696 &    0.511557 &  0.530779 &  0.457785 &  0.208321 &  0.439915 &  0.405056 &  0.516199 \\
kylin &  0.316579 &    0.448951 & \cellcolor[HTML]{e7e7e7} 0.466410 &  0.322363 &  0.111994 &  0.435269 &  0.332273 &  0.411770 \\
lens &  0.374882 &    0.480496 & \cellcolor[HTML]{e7e7e7} 0.559902 &  0.328536 &  0.131715 &  0.465462 &  0.408012 &  0.470915 \\
mahout &  0.482755 &    0.636575 & \cellcolor[HTML]{e7e7e7} 0.636699 &  0.463299 &  0.153698 &  0.613597 &  0.280598 &  0.585866 \\
manifoldcf &  0.463938 &    0.534746 & \cellcolor[HTML]{e7e7e7} 0.659084 &  0.578284 &  0.288716 &  0.606849 &  0.650394 &  0.575035 \\
nutch &  0.442079 &    0.510856 &  0.508616 &  0.412224 &  0.196293 & \cellcolor[HTML]{e7e7e7} 0.512133 &  0.430201 &  0.495843 \\
opennlp &  0.479977 &    0.541031 &  0.529440 &  0.430390 &  0.166851 & \cellcolor[HTML]{e7e7e7} 0.577043 &  0.382945 &  0.566229 \\
parquet-mr &  0.516109 & \cellcolor[HTML]{e7e7e7}   0.645096 &  0.612779 &  0.442167 &  0.135967 &  0.613087 &  0.259887 &  0.504891 \\
santuario-java &  0.454974 &    0.597219 & \cellcolor[HTML]{e7e7e7} 0.653833 &  0.464933 &  0.209318 &  0.602171 &  0.472031 &  0.651655 \\
systemml &  0.349762 &    0.342008 & \cellcolor[HTML]{e7e7e7} 0.386711 &  0.293698 &  0.113281 &  0.305690 &  0.354479 &  0.355639 \\
tika &  0.513870 &    0.616276 & \cellcolor[HTML]{e7e7e7} 0.639042 &  0.558498 &  0.298444 &  0.581097 &  0.376991 &  0.590818 \\
wss4j &  0.228257 &    0.323039 & \cellcolor[HTML]{e7e7e7} 0.341043 &  0.192375 &  0.084688 &  0.327874 &  0.117347 &  0.285987 \\
\midrule
mean value  & 0.418$\pm$0.098 &  0.505$\pm$0.109 & 0.542$\pm$0.111 &  0.390$\pm$0.107 & 0.164$\pm$0.066 & 0.513$\pm$0.118 & 0.337$\pm$0.111 & 0.517$\pm$0.116 \\
standard deviation & 0.129 & 0.143 & 0.145 & 0.140 & 0.087 & 0.155 & 0.145 & 0.152 \\
\bottomrule
\end{tabular}%
}
\caption{\gls{MRR} values of the different bug localization approaches using the SmartSHARK data set}
\label{tab:mrr_experiment2}
\end{table}

\begin{table}[]
\centering
\resizebox{\textwidth}{!}{%
\begin{tabular}{lrr|rr}
\toprule
Project &  MAP\_MultiVersion &  MAP\_Time-Aware &  MRR\_MultiVersion &  MRR\_Time-Aware \\
\midrule
ant-ivy  &      0.496122 &   \cellcolor[HTML]{e7e7e7}    0.502094 &      0.538422 &    \cellcolor[HTML]{e7e7e7}   0.548233 \\
archiva  &      0.370623 &    \cellcolor[HTML]{e7e7e7}   0.413654 &      0.408797 &    \cellcolor[HTML]{e7e7e7}   0.459270 \\
calcite  &      0.561968 &   \cellcolor[HTML]{e7e7e7}    0.574669 &      0.615734 &   \cellcolor[HTML]{e7e7e7}    0.628116 \\
cayenne  &  \cellcolor[HTML]{e7e7e7}    0.402604 &       0.401408 &      0.425274 &    \cellcolor[HTML]{e7e7e7}   0.425835 \\
commons-bcel  &   \cellcolor[HTML]{e7e7e7}   0.425496 &    \cellcolor[HTML]{e7e7e7}   0.425496 &  \cellcolor[HTML]{e7e7e7}    0.436706 &   \cellcolor[HTML]{e7e7e7}    0.436706 \\
commons-beanutils  &    \cellcolor[HTML]{e7e7e7}  0.704911 &     \cellcolor[HTML]{e7e7e7}  0.704911 &   \cellcolor[HTML]{e7e7e7}   0.765278 &  \cellcolor[HTML]{e7e7e7}     0.765278 \\
commons-codec  &      0.736462 &   \cellcolor[HTML]{e7e7e7}    0.743128 &      0.838889 &    \cellcolor[HTML]{e7e7e7}   0.845556 \\
commons-collections  &      0.268564 &  \cellcolor[HTML]{e7e7e7}     0.770596 &      0.289431 &    \cellcolor[HTML]{e7e7e7}   0.817886 \\
commons-compress   &      0.659129 &   \cellcolor[HTML]{e7e7e7}    0.667731 &      0.737325 &    \cellcolor[HTML]{e7e7e7}   0.742715 \\
commons-configuration   &      0.576907 &   \cellcolor[HTML]{e7e7e7}    0.651290 &      0.659275 &    \cellcolor[HTML]{e7e7e7}   0.743606 \\
commons-dbcp &      0.399022 &   \cellcolor[HTML]{e7e7e7}    0.643933 &      0.428291 &    \cellcolor[HTML]{e7e7e7}   0.682051 \\
commons-digester &  \cellcolor[HTML]{e7e7e7}    0.605556 &   \cellcolor[HTML]{e7e7e7}    0.605556 &   \cellcolor[HTML]{e7e7e7}   0.700000 &    \cellcolor[HTML]{e7e7e7}   0.700000 \\
commons-io &    \cellcolor[HTML]{e7e7e7}  0.825370 &       0.823519 &  \cellcolor[HTML]{e7e7e7}    0.894444 &       0.892593 \\
commons-jcs &      0.347422 &    \cellcolor[HTML]{e7e7e7}   0.616191 &      0.375926 &     \cellcolor[HTML]{e7e7e7}  0.656790 \\
commons-jexl &      0.387787 &  \cellcolor[HTML]{e7e7e7}     0.551408 &      0.436508 &   \cellcolor[HTML]{e7e7e7}    0.616631 \\
commons-lang &      0.702518 &  \cellcolor[HTML]{e7e7e7}     0.769651 &      0.786547 &   \cellcolor[HTML]{e7e7e7}    0.865801 \\
commons-math &      0.451436 &    \cellcolor[HTML]{e7e7e7}   0.669190 &      0.504254 &   \cellcolor[HTML]{e7e7e7}    0.739494 \\
commons-net &      0.774366 &    \cellcolor[HTML]{e7e7e7}   0.788611 &      0.818600 &    \cellcolor[HTML]{e7e7e7}   0.835694 \\
commons-scxml &    \cellcolor[HTML]{e7e7e7}  0.666737 &       0.661835 &    \cellcolor[HTML]{e7e7e7}  0.708333 &       0.703431 \\
commons-validator &  \cellcolor[HTML]{e7e7e7}    0.738481 &       0.726629 &   \cellcolor[HTML]{e7e7e7}   0.792275 &       0.771905 \\
commons-vfs &      0.374173 &   \cellcolor[HTML]{e7e7e7}    0.650723 &      0.388341 &    \cellcolor[HTML]{e7e7e7}   0.669307 \\
deltaspike &      0.498238 &    \cellcolor[HTML]{e7e7e7}   0.528142 &      0.530922 &    \cellcolor[HTML]{e7e7e7}   0.563055 \\
eagle &      0.349435 &    \cellcolor[HTML]{e7e7e7}   0.368300 &      0.376595 &     \cellcolor[HTML]{e7e7e7}  0.394984 \\
giraph &      0.473961 &   \cellcolor[HTML]{e7e7e7}    0.586550 &      0.499654 &    \cellcolor[HTML]{e7e7e7}   0.617697 \\
gora &      0.610263 &    \cellcolor[HTML]{e7e7e7}   0.651899 &      0.630034 &   \cellcolor[HTML]{e7e7e7}    0.685917 \\
jspwiki &      0.122252 &     \cellcolor[HTML]{e7e7e7}  0.305153 &      0.133227 &   \cellcolor[HTML]{e7e7e7}    0.355245 \\
knox &      0.473973 &    \cellcolor[HTML]{e7e7e7}   0.475227 &      0.516199 &    \cellcolor[HTML]{e7e7e7}   0.516423 \\
kylin &      0.396795 &   \cellcolor[HTML]{e7e7e7}    0.459309 &      0.411770 &   \cellcolor[HTML]{e7e7e7}    0.477426 \\
lens &   \cellcolor[HTML]{e7e7e7}   0.416356 &       0.415731 &      0.470915 &    \cellcolor[HTML]{e7e7e7}   0.472643 \\
mahout &      0.556360 &    \cellcolor[HTML]{e7e7e7}   0.685342 &      0.585866 &   \cellcolor[HTML]{e7e7e7}    0.727983 \\
manifoldcf &      0.547109 &  \cellcolor[HTML]{e7e7e7}     0.566626 &      0.575035 &    \cellcolor[HTML]{e7e7e7}   0.595395 \\
nutch &      0.483704 &     \cellcolor[HTML]{e7e7e7}  0.595502 &      0.495843 &    \cellcolor[HTML]{e7e7e7}   0.609214 \\
opennlp &      0.546073 &   \cellcolor[HTML]{e7e7e7}    0.546456 &      0.566229 &    \cellcolor[HTML]{e7e7e7}   0.566612 \\
parquet-mr &      0.461694 &     \cellcolor[HTML]{e7e7e7}  0.575407 &      0.504891 &     \cellcolor[HTML]{e7e7e7}  0.642112 \\
santuario-java &      0.622546 &    \cellcolor[HTML]{e7e7e7}   0.735722 &      0.651655 &   \cellcolor[HTML]{e7e7e7}    0.775557 \\
systemml &      0.340754 &   \cellcolor[HTML]{e7e7e7}    0.347954 &      0.355639 &    \cellcolor[HTML]{e7e7e7}   0.362745 \\
tika &      0.543742 &   \cellcolor[HTML]{e7e7e7}    0.558409 &      0.590818 &   \cellcolor[HTML]{e7e7e7}    0.605947 \\
wss4j &      0.268095 &    \cellcolor[HTML]{e7e7e7}   0.541335 &      0.285987 & \cellcolor[HTML]{e7e7e7}      0.587095 \\
\midrule 
mean value  & 0.505$\pm$0.080 &  0.587$\pm$0.067  &  0.546$\pm$0.089 & 0.634$\pm$0.074 \\
standard deviation & 0.156 &   0.131 &    0.173 & 0.144  \\
\bottomrule
\end{tabular}%
}
\caption{\gls{MAP} and \gls{MRR} values using Broccoli on Time-Aware matching and multi version matching the SmartSHARK data set}
\label{tab:project_timeaware}
\end{table}

\end{document}